\documentclass[hyper,12pt]{JHEP3}
\usepackage{epsfig}
\usepackage{amsmath}
\usepackage{amssymb}
\usepackage{epsf}
\usepackage{graphicx}
\usepackage{cite}
\usepackage{graphics}
\usepackage{epstopdf}

\font\cmr=cmr7

\title{Hard pion and prompt photon at RHIC, \\ from single to double inclusive production}

\author{
Fran\c{c}ois Arleo\\~\\
CERN, PH-TH Dept. \\1211 Geneva 23, Switzerland\\~\\
E-mail: \email{arleo@cern.ch}
}

\abstract{
Single pion and prompt photon large transverse momentum spectra in $p$--$p$ and Au--Au collisions are computed in perturbative QCD at RHIC energy, $\sqrt{s_{_{{\rm N N}}}} = 200$~GeV. Next-to-leading order calculations are discussed and compared with $p$--$p$ scattering data. Subsequently, quenching factors are computed to leading order for both pions and photons within the same energy loss model. The good agreement with PHENIX preliminary data allows for a lower estimate of the energy density reached in central Au--Au collisions, $\epsilon_{_{\rm RHIC}}\gtrsim~10$~GeV/fm$^3$. Double inclusive $\gampi$ production in $p$--$p$ and Au--Au collisions is then addressed. Next-to-leading order corrections prove rather small in $p$--$p$ scattering. In Au--Au collisions, the quenching of momentum-correlation spectra is seen to be sensitive to parton energy loss processes, which would help to understand how the fragmentation dynamics is modified in nuclear collisions at RHIC.}

\keywords{QCD, Jets, Hadronic Colliders}

\preprint{CERN-PH-TH/2005-222\\hep-ph/0601075}

\newcommand{\dd}{{\rm d}\,}

\newcommand{\dsigd}{  {\dd \sigma^{\gamma \,{\rm dir}} 
                \over {\dd {\bf  p_{_\perp}} \dd y}}}
\newcommand{\dsigb}{  {\dd \sigma^{{\gamma \, \rm frag}} 
                \over {\dd {\bf p_{_\perp}} \dd y}}}
\newcommand{\dsigij}{  {\dd {\widehat \sigma}_{ij} 
                \over {\dd {\bf p_{_\perp}} \dd y}}}
\newcommand{\dsigijk}{  {\dd {\widehat \sigma}_{ij}^k 
                \over {\dd {\bf p_{_\perp}} \dd y}}}
\newcommand{\kd}{  K^{\hbox{\cmr dir}} }
\newcommand{\kb}{  K^{\hbox{\cmr frag}} }

\newcommand{\alfspi}{ {\alpha_s(\mu) \over 2 \pi} }

\newcommand{\mD}{ {m_{_D}} }
\newcommand{\alfas}{ {\alpha_s} }
\newcommand{\alfasd}{ {\alpha^2_s} }
 
\newcommand{\be}{\begin{equation}} \newcommand{\ee}{\end{equation}}
\newcommand{\bea}{\begin{eqnarray}} \newcommand{\ena}{\end{eqnarray}}

\def\cO#1{{{\cal{O}}}\left(#1\right)} \def\pt{$p_{_T}$}
\def\ptpi{p_{_{\perp_\pi}}} \def\ptgamma{p_{_{\perp_\gamma}}}
\def\m{m_{_{\gamma \pi}}}
\def\qt{q_{_{\perp_{\gamma \pi}}}}
\def\kt{{k_{_\perp}}} \def\pt3{{p_{_T{_3}}}}
\def\gampi{\gamma$--$\pi^0\ } 
\def\cO#1{{{\cal{O}}}\left(#1\right)} \def\pt{$p_{_T}$}
\def\pt{p_{_\perp}}
\def\picut{p_{_{\perp_\pi}}^{\rm cut}}
\def\gacut{p_{_{\perp_\gamma}}^{\rm cut}}
\def\phigampi{\phi_{_{\gamma \pi}}}
 
\def\X{{\rm X}}
\def\z{z_{_{\gamma \pi}}}
\def\zstar{z^*_{_{\gamma, \pi}}}
\def\zz{z_{_{\gamma, \pi}}}
\def\muopt{\mu_{_{{\rm opt}}}}
\def\muoptpi{\mu_{_{{\rm opt}}}^\pi}
\def\muoptga{\mu_{_{{\rm opt}}}^\gamma}
\def\gamgam{\gamma-\gamma\ }
\def\gampi{\gamma$--$\pi^0\ }

\begin{document}

\section{Introduction}

Photon production is a promising observable in high energy heavy-ion collisions. On the one hand, one expects that hot media such as quark--gluon plasma radiate thermal photons~\cite{Feinberg:1976ua,Shuryak:1978ij,Aggarwal:2000th,Arleo:2003gn} with transverse momenta of the order of its temperature, $\ptgamma = \cO{T}$. The huge decay background from neutral pions, especially below 4~GeV, makes the experimental extraction of such a signal quite a difficult task. In that respect, the precise PHENIX measurements recently shown at Quark Matter down to 1~GeV are remarkable~\cite{Akiba:2005pc,Bathe:2005pc}. What is more, the possible excess in photon production reported in the 1--3~GeV range above the expected QCD rate is exciting; yet any definite conclusion would be highly premature. On the other hand and side of the spectrum, the hard prompt photon signal with $\ptgamma \gg T$ can be used to gauge the strength of nuclear effects in hard processes (see e.g. Refs.~\cite{Owens:1986mp,Aurenche:1987fs,Aurenche:1992yc,Gordon:1993qc,Aurenche:1998gv,Binoth:1999qq,Aurenche:1999nz} on prompt photon phenomenology). In particular, comparing blind hard probes --~Drell--Yan, heavy bosons, prompt photons~-- with coloured hard probes --~hadrons, jets~-- would allow for a clear experimental distinction between initial-state effects (such as small-$x$ saturation physics~\cite{Iancu:2003xm}) and final-state interactions (e.g. parton energy loss processes~\cite{Baier:2000mf,Gyulassy:2003mc}). However, it may be misleading to assume that prompt photon production, because of its colour neutrality, should not depend on the dense-medium properties. Indeed, to leading order in the perturbative expansion, prompt photons can be produced directly in the hard subprocess (``Drell--Yan-like'') but also from the collinear fragmentation of a hard quark or gluon (``jet-like'')~\cite{Aurenche:1998gv}. Of course, only the sum of these two components is meaningful and as scale-independent as possible: the leading-order (LO) fragmentation process may be seen as a next-to-leading order (NLO) direct contribution and vice~versa, depending on the resolution scale. Nevertheless, it is worth stressing that the prompt photon signal could in principle be modified by the dense-medium produced in the nucleus--nucleus reaction. In particular, the mechanism that spectacularly quenches the pion $\ptpi$ spectra in Au--Au collisions~\cite{Adcox:2001jp,Adler:2002xw,Adler:2003qi} should also affect prompt photon production, although not necessarily with a similar strength. It is thus important to treat on an equal footing large-$\pt$ pion and prompt photon production, from $p$--$p$ to A--A collisions.

Single hadron spectra are certainly useful to reveal the formation of a dense medium in heavy-ion collisions (as we shall see, the present data actually allow for a lower estimate of the energy density reached in central Au--Au collisions), however such measurements hardly inform us on how exactly the medium affects fragmentation functions and, more generally, on the fragmentation process itself. Indeed, the initial  parton momentum $k_{_{\perp_i}}$ --~hence the fragmentation variable $z = p_{_{\perp_h}} / k_{_{\perp_i}}$ entering fragmentation functions~-- is not fixed. As a consequence, there is a clear need to go beyond the single particle production picture. This triggered in particular several measurements~\cite{Adler:2002tq,Adams:2004wz,Dietel:2005st,Chiu:2002ma,Adler:2005ee} and calculations~\cite{Wang:2003mm,Qiu:2003pm} of 2-particle azimuthal correlations. Moreover, following Ref.~\cite{Wang:1996yh}, performing photon-tagged {\it momentum} correlations in the double inclusive production of $\gampi$ pairs at the LHC has been proposed as a powerful tool to extract (or at least to constrain) fragmentation functions: to leading order, the hard photon (hopefully produced directly) gives access to the leading {\it parton} transverse momentum, which eventually fragments into the pion~\cite{Arleo:2004xj}. The wealth of RHIC large-$\pt$ data provides hints that the energy loss process is probably at work in Au--Au central collisions~\cite{Adcox:2004mh,Adams:2005dq,Eskola:2005tx} (and to a lesser extent in Cu--Cu~\cite{Shimomura:2005pc}), such a picture being supported by many phenomenological analyses~\cite{Vitev:2002pf,Eskola:2004cr,Dainese:2004te,Turbide:2005fk,Borghini:2005em}. Consequently, addressing such $\gampi$ momentum correlations also at RHIC energy is particularly relevant.

In this paper, we first explore parton energy loss effects on both single hard pion and single prompt photon production within the same model, based on the Baier--Dokshitzer--Mueller--Peign\'e--Schiff (BDMPS) framework~\cite{Baier:1997kr,Baier:1997sk}. After discussing single spectra in $p$--$p$ collisions computed in QCD at NLO, we predict the expected pion and photon quenching factors and compare them with the available data. The photon total           yield over the background is investigated as well. A lower estimate for the RHIC energy density is then determined. The second part of this work is devoted to the study of photon-tagged momentum correlations. Various absolute correlation spectra are constructed in $p$--$p$ and their expected quenching in central Au--Au collisions is discussed. Counting rates are also given. Let us now start with the details of the perturbative calculations.

\section{Preamble: predictions}
\label{sec:predictions}

We discuss in this section the accuracy as well as the ingredients used in the present perturbative QCD predictions of single and double inclusive pion and photon production in $p$--$p$ and Au--Au collisions.

\subsection{Proton--proton collisions}
\label{pp}

\subsubsection{Single inclusive production}
\label{sec:ppsingle}

Single inclusive pion and photon hadroproduction cross sections in $p$--$p$ collisions are computed at NLO accuracy in QCD, using the work of Ref.~\cite{Aurenche:1998gv}.  For single-pion production, NLO cross sections read~\cite{Aurenche:1998gv}
\begin{eqnarray}
\label{eq:pisingle}
{\dd \sigma^{\pi} \over \dd {\bf p_\perp} \dd y} &=& \sum_{i,j,k=q,g} \int \dd x_1 \dd x_2 
F_{i/p}(x_1, M) F_{j/p}(x_2,M) {\dd z \over z^2} D_{\pi/k}(z,M_F) 
\nonumber \\
&&{}\times \left [\left ( {\alpha_s (\mu ) \over 2 \pi} \right )^2
{\dd \widehat{\sigma}_{ij}^{k} \over \dd {\bf p_\perp} \dd y}
+ \left ( {\alpha_s(\mu) \over 2 \pi} \right )^3 K_{ij,k}(\mu , M, M_F)
\right ],
\end{eqnarray}
where $F_{i,j/p}$ are the proton parton distribution functions (PDF), $D_{\pi/k}$ the pion fragmentation functions, and $\widehat{\sigma}_{ij}^{k}$ (respectively $K_{ij,k}$) the leading-order (respectively next-to-leading order) partonic cross section in the $\overline{{\rm MS}}$ scheme\footnote{We omit the explicit dependence of $\widehat{\sigma}_{ij}^{k}$ and $K_{ij,k}$ on the kinematic variables $x_1$, $x_2$, ${\sqrt s}$, $\pt$, and $y$ for clarity.}. We denote by $\mu$, $M$ and $M_{F}$ the renormalization, the factorization and the fragmentation scale.

Unlike pions, single prompt photons can be produced ``directly'' in the hard subprocess~\cite{Aurenche:1998gv}, with a contribution
\bea
\dsigd &=& \sum_{i,j=q,g} \int \dd x_{1} \dd x_{2}
\ F_{i/p}(x_{1},M)\ F_{j/p}(x_{2},M) \nonumber \\
&\ & \qquad \qquad
\alfspi \left( \dsigij \ + \ \alfspi \kd_{ij} (\mu,M,M_{_F}) \right),
\label{eq:dir}
\ena
in addition to the collinear fragmentation process~\cite{Aurenche:1998gv}:
\bea
\dsigb &=& \sum_{i,j,k=q,g} \int \dd x_{1} \dd x_{2}{ \dd z \over z^2}
\ F_{i/p}(x_{1},M)\ F_{j/p}(x_{2},M)\ D_{\gamma/k}(z, M_{_F})
\nonumber \\
&\ & \qquad \qquad
\ \left( \alfspi \right)^2\  \left( \dsigijk \ + 
\ \alfspi \kb_{ij,k} (\mu,M,M_{_F}) \right),
\label{eq:brem}
\ena 
at the same order in the perturbative expansion. Let us once more emphasize that the distinction between direct and fragmentation photons is arbitrary, only the sum of these two contributions being meaningful and with a lesser fragmentation scale dependence~\cite{Aurenche:1998gv}.

\subsubsection{Double inclusive production}
\label{sec:ppdouble}

At leading order in QCD, the basic two-particle $\gampi$ correlation cross section, from which various observables are constructed, can be written~\cite{Chiappetta:1996wp,Binoth:2002wa}:
\begin{eqnarray}                                                                
{\dd \sigma^{^{pp \rightarrow \gamma\pi}} \over \dd p_{_\perp{_\pi}} \dd y_\pi \dd z_\pi \dd                 
p_{_\perp{_\gamma}} \dd y_\gamma \dd z_\gamma} = {1 \over 8 \pi s^2} & \sum_{a,b,c,d} &  {               
D_{\pi/c}(z_\pi,M_{_F})\over z_\pi } { D_{\gamma/d}(z_\gamma,M_{_F})\over z_\gamma }\                
k_{_{\perp{_c}}} \  \delta(k_{_{\perp{_c}}}-k_{_{\perp{_d}}}) \nonumber\\  &&                     
{F_{{a/p}}(x_1,M) \over x_1} \ { F_{{b/p}}(x_2,M) \over x_2},  \ \               
|{\overline {\cal M}}|^2_{ab \rightarrow cd}                                           
\label{eq:correl-rate}                                                          
\end{eqnarray}
where ${\cal M}$ is the LO hard $a\,b\,\to\,c\,d$ scattering amplitude and the distinction between the direct and the fragmentation component is made implicit in the $D_{\gamma/d}(z_\gamma,M_{_F})$ fragmentation functions.

\subsection{Nucleus--nucleus collisions}
\label{sec:aa}

The status of NLO QCD calculations in nucleus--nucleus collisions is not as yet well established. Therefore, we shall consider LO calculations and only show the normalized ratio of the Au--Au over the $p$--$p$ production cross section:
\begin{equation}
R(p_{_\perp}) = \frac{1}{N_{_{\rm {coll}}}} \,\, \frac{\sigma_{_{{\rm NN}}}}{\sigma^{{\rm geo}}_{_{{\rm Au Au}}}} \,\times\, {d\sigma_{_{\rm {Au~Au}}}^{\gamma \pi} \over d{\bf p_\perp} \dd y} \biggr/  {d\sigma_{_{p p}}^{\gamma \pi} \over d{\bf p_\perp} \dd y},
\end{equation}
where $\sigma^{{\rm geo}}_{_{{\rm Au Au}}}$ is the geometric cross section obtained via the Glauber multiple scattering theory, $\sigma_{_{{\rm NN}}}$ the nucleon--nucleon cross section, and $\langle N_{{\rm coll}} \rangle \big |_{_{\cal C}}$ the number of binary collisions in a given centrality class ${\cal C}$. Numerically, we have $\sigma^{{\rm geo}}_{_{{\rm AA}}} = 6900$~mb, $\sigma_{_{{\rm NN}}} = 42$~mb and $\langle N_{{\rm coll}}\rangle\!|_{\cal C} =  779$ in central (${\cal C} \le 20\%$) Au--Au collisions at RHIC energy~\cite{Arleo:2003gn,Adler:2003qi}. 

Ignoring any nuclear effect, the nucleus--nucleus collision cross section is deduced directly from the $p$--$p$ LO cross section, replacing the proton PDF in Eqs.~(\ref{eq:pisingle}) to (\ref{eq:correl-rate}) by
\begin{equation}\label{eq:iso}
F_{i/A}(x,M) = Z \, F_{i/p}(x,M) + (A-Z) \, F_{i/n}(x,M),
\end{equation}
where $Z$ and $A$ are respectively the number of protons and the atomic mass number of each nucleus. The neutron PDF $F_{i/n}$ in Eq.~(\ref{eq:iso}) is obtained from the proton $F_{i/p}$ by the usual isospin conjugation assumptions: $u^p = d^n$, $d^p = u^n$, $\bar{u}^p = 
\bar{d}^n$, $\bar{d}^p = \bar{u}^n$, and $\bar{s}^p = \bar{s}^n$. 

Such possible isospin effects, when comparing different nuclear targets --~and in particular almost isoscalar nuclei ($A \simeq 2\,Z$) such as Au with a proton ($A = Z$)~-- may be significant when hadron or photon production occurs at large Bjorken $x=~\cO{2 p_{_\perp}/\sqrt{s}}\lesssim 1$, at which the partonic process involves essentially the scattering of valence quarks. Conversely, it should remain completely negligible, say, around~$x\simeq~0.01$, below which the nucleon PDF is dominated by the gluons. Note that such an effect is of course strongly magnified in electromagnetic processes such as prompt photon production because of the valence quark electric charges, as we shall see in Section~\ref{sec:aasingle}.

\subsubsection{Shadowing}
\label{sec:shadowing}

On top of these isospin corrections, parton densities are known to be modified in a nuclear environment over the whole Bjorken-$x$ range (see~\cite{Piller:1999wx,Arneodo:1994wf,Accardi:2003be} for reviews). To take into account such shadowing effects, we use the global LO QCD fit of Deep Inelastic Scattering (DIS) and Drell--Yan data performed by Eskola, Kolhinen, and Salgado (EKS98)~\cite{Eskola:1998df}, who extract the ratio $S_{a/{\rm A}}(x,M)$ of the nuclear PDF over the free proton one:
\begin{equation}\label{eq:eks}
  S_{a/{\rm A}}(x,M) = \frac{F_{a/{\rm A}}(x,M)}{A\,F_{a/p}(x,M)}.
\end{equation}
The ratio~(\ref{eq:eks}) depends on each parton species $a$ as well as on the factorization scale $M$ through DGLAP evolution. The function $S_{a/{\rm A}}$ is smaller than 1 at small $x \ll 0.01$ (shadowing) and large $x \gtrsim 0.3$ (EMC effect), while slightly larger than 1 (antishadowing) for valence quarks and gluons at $x\simeq~0.1$. Note that the small-$x$ region is only poorly constrained by the currently available data, hence quite uncertain~\cite{Armesto:2003bq}.

In the calculations to come, we shall mark the difference between our predictions in Au--Au collisions when shadowing effects are taken into account (labelled ``Au~Au EKS98'' in the figures) and where they are not (``Au~Au'').

\subsubsection{Energy loss}
\label{sec:energy_loss}

In the nuclear predictions discussed so far, we assumed implicitly that the hot and dense-medium probably produced in high energy heavy-ion collisions does not modify either the pion or the prompt photon production process. 

We now suppose that the quarks and gluons produced in the partonic subprocess with momentum $k_{_\perp}$ undergo multiple scattering in the medium. Doing so, they lose an amount of energy $\epsilon$ with a probability ${\cal P}(\epsilon, k_{_\perp})$. Using a Poisson approximation for the soft gluon emission process, this probability distribution is related to the medium-induced gluon spectrum $dI/d\omega$ in~\cite{Baier:2001yt} and later determined explicitly in Refs.~\cite{Arleo:2002kh,Salgado:2002cd}. In order to make the connection between the energy loss process and the quenching finally observed, we follow here the approach of Ref.~\cite{Wang:1996yh}, in which the fragmentation variable is shifted from $z$ to $z^*=z/(1-\epsilon/\kt)$. The medium-modified fragmentation functions in that model thus read~\cite{Wang:1996yh}
\begin{equation}
  \label{eq:modelFF}
\zz\,D_{\gamma,\pi/d}^{\rm med}(\zz, M_{_F}, \kt) = \int_0^{\kt (1
  - \z)} \, \dd \epsilon \,\,{\cal P}_d(\epsilon, \kt)\,\,\,
\zstar\,D_{\gamma,\pi/d}(\zstar, M_{_F}).
\end{equation}
Let us mention that another attempt at modelling fragmentation functions in the medium, accounting for all leading and subleading successive parton branchings, has been performed recently within the modified leading-logarithmic approximation~\cite{Borghini:2005em}.

The typical amount of energy loss depends on the one scale entering the medium-induced gluon spectrum, $\omega_c = 1/2\,\,\hat{q}\,\,L^2$; here the transport coefficient $\hat{q}$ is defined as the typical kick in transverse momentum space per unit length that the hard gluons undergo, and $L$ is the medium path length. While $L$ should be integrated over the whole production volume, we shall take here for simplicity a mean length $\langle L \rangle = 5$~fm in the following calculations. Such a model was shown to describe successfully the observed hadron attenuation measured by EMC~\cite{Ashman:1991cx} and HERMES~\cite{Airapetian:2003mi} in semi-inclusive DIS on nuclear targets~\cite{Arleo:2003jz}.

Although perturbative estimates for the transport coefficient in a hot quark--gluon plasma have been suggested~\cite{Baier:1997sk}, the precise value of $\omega_c$ for the medium produced at RHIC energy is somewhat difficult to estimate from first principles. Based on the quenching of single-inclusive pion data measured by PHENIX in central Au--Au collisions (discussed in Section~\ref{sec:aasingle}), we shall use the $\omega_c = 20$--$25$~GeV range in our calculations. This estimate will be critically discussed in Section~\ref{sec:rhicdensity}.

\subsection{Ingredients}
\label{sec:ingredients}

The proton parton distribution functions $F_{i,j/p}$ are taken from the NLO (LO) CTEQ6M (CTEQ6L) parametrization~\cite{Pumplin:2002vw}. The Kniehl--Kramer--P\"otter (KKP) (N)LO fragmentation functions into neutral pions~\cite{Kniehl:2000fe} were used. The Bourhis--Fontannaz--Guillet--Werlen (BFGW)~\cite{Bourhis:1997yu,Bourhis:2000gs} NLO fragmentation functions for photons were chosen for both LO and NLO computation in view of the lack of recent leading-order determinations of photon fragmentation functions. All scales were taken to be equal, $\mu = M = M_F$, and chosen so as to minimize the scale-dependence of the NLO predictions.  In order to investigate the scale sensitivity of the single-inclusive QCD calculations, hence part of the theoretical uncertainty\footnote{Important uncertainties also arise from the poorly known fragmentation functions.}, all scales were allowed to vary simultaneously in a given range around the optimal scale. The scale-fixing procedure in single-inclusive particle production will be discussed in Section~\ref{sec:scalefixing}. In the double-inclusive $\gampi$ channel, all scales are taken to be given by half the prompt photon momentum.

\section{Single inclusive pion and photon production}
\label{sec:single}

\subsection{Scale-fixing procedure}
\label{sec:scalefixing}

Before comparing NLO pQCD calculations with RHIC data, all scales --~which should be $\cO{\pt}$~-- have to be fixed within a given prescription. Let $\muopt$ be the optimal scale, defined as the one which minimizes the scale-dependence of the NLO predictions:
\begin{equation}\label{eq:opt}
\frac{\partial}{\partial \mu}\, \frac{\dd \sigma}{\dd p_{_{\perp}}}  \Bigg |_{\muopt} \simeq 0.
\end{equation}
Although $\muopt / \pt$ may in principle depend on $\pt$, it is fixed here in an arbitrary (but sufficiently hard) $\pt = 20$~GeV value. 

\begin{figure}[ht]
  \begin{minipage}[ht]{7.8cm}
    \begin{center}
      \includegraphics[height=7.8cm]{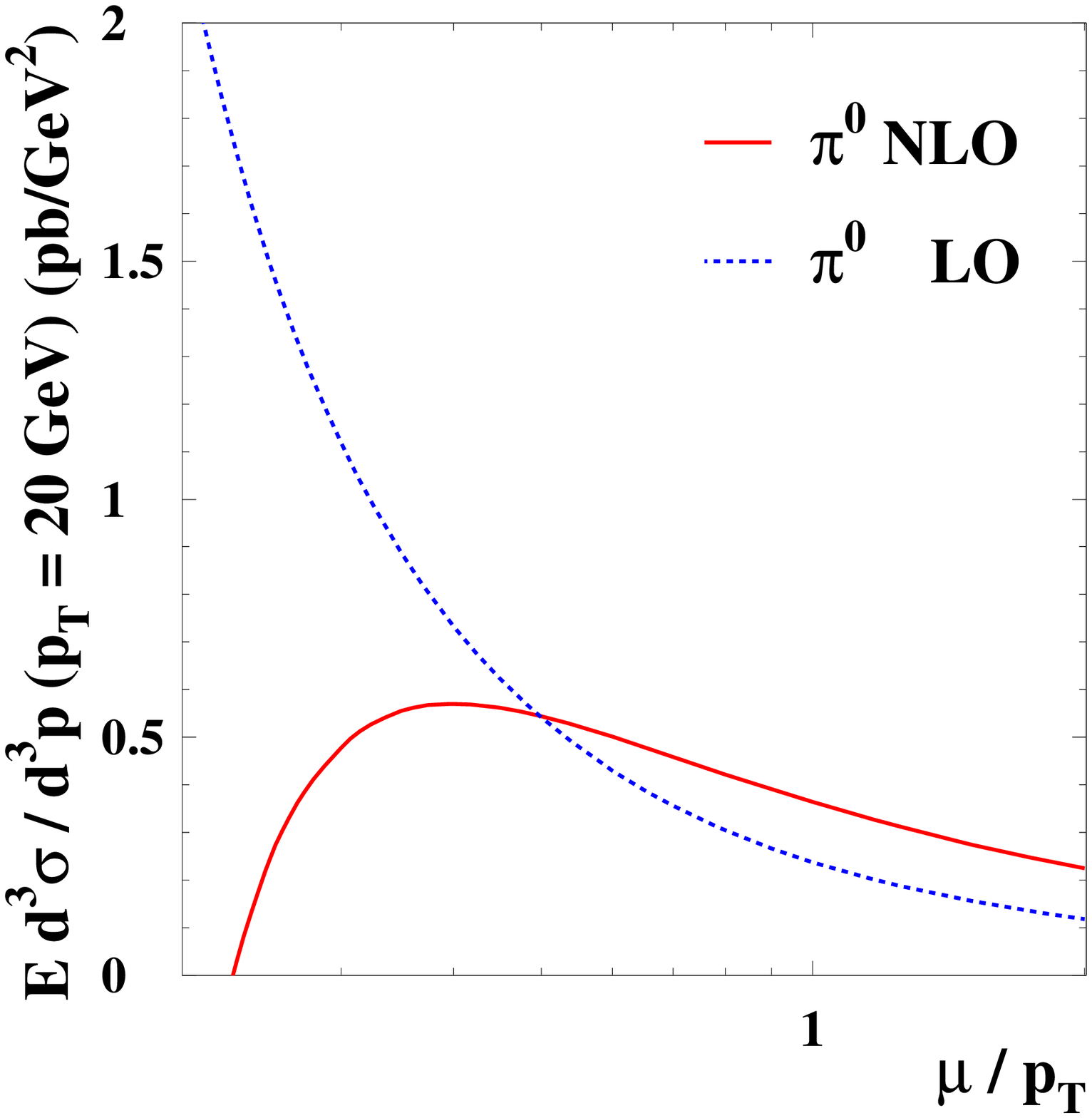}
    \end{center}
  \end{minipage}
~
  \begin{minipage}[ht]{7.8cm}
    \begin{center}
      \includegraphics[height=7.8cm]{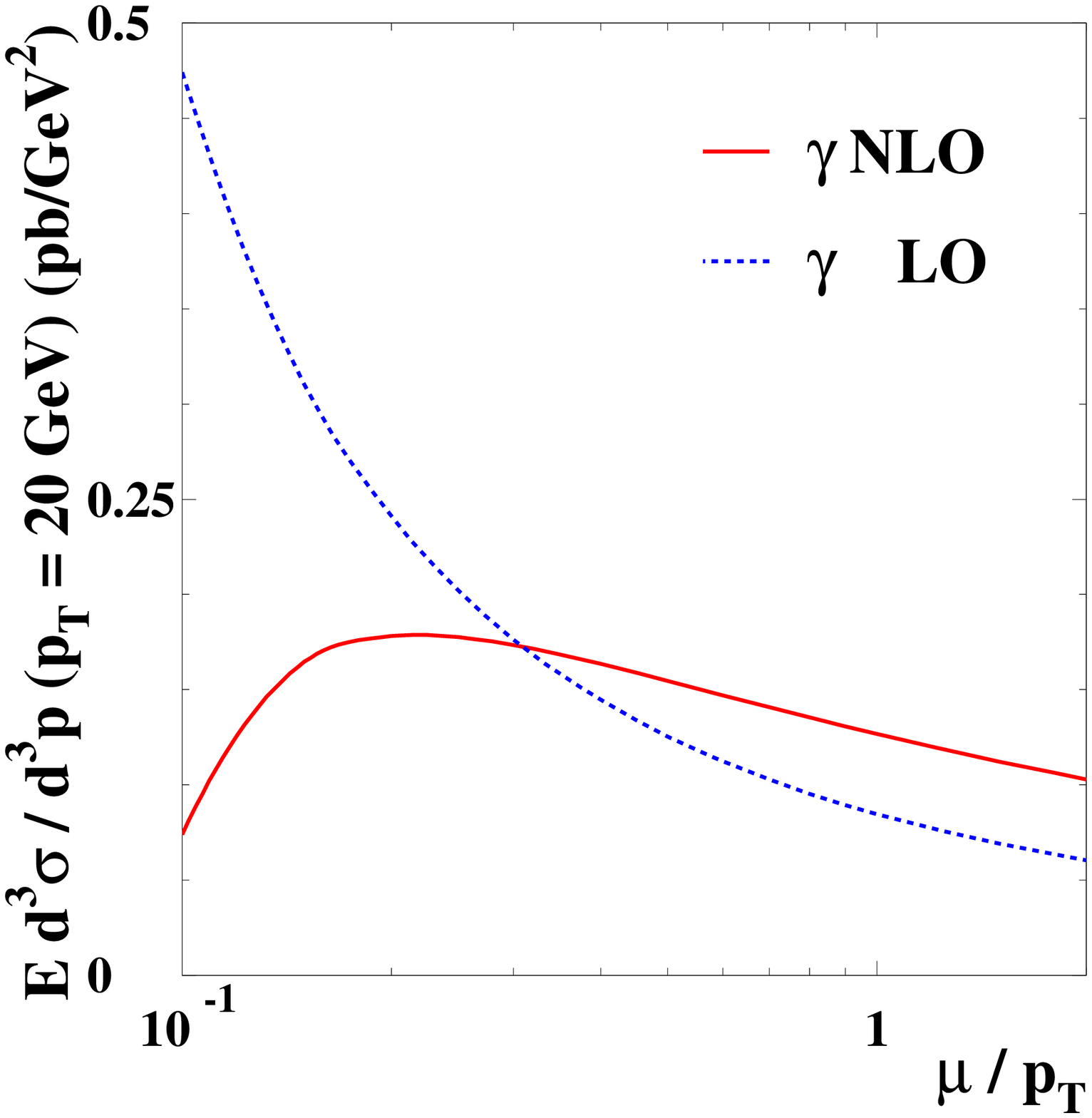}
    \end{center}
  \end{minipage}
  \caption{LO (dashed) and NLO (solid) predictions for pions (left) and prompt photons (right) at $\pt = 20$~GeV are shown as a function of the renormalization scale, $\mu$ (normalized to $\pt$). The factorization scale $M$ and the fragmentation scale $M_F$ are taken to be equal to $\mu$. The optimal scale is chosen so as to minimize the scale dependence of the NLO predictions (see text).}
  \label{fig:opt}
\end{figure}

In Fig.~\ref{fig:opt} it is shown the scale variation of the LO (dashed) and NLO (solid) predictions for pions (left) and prompt photons (right). As expected, NLO predictions prove to be much more stable than the LO calculations. In particular, a stable point can be found at NLO accuracy while the LO cross sections monotonically decrease with the scale $\mu$. From Fig.~\ref{fig:opt}, the optimal scale defined in Eq.~(\ref{eq:opt}) proves to be actually well below $\pt$: roughly $\muoptpi \simeq \pt / 2$ for pions and $\muoptga = \pt / 4$ for photons. Since we want to perform perturbative calculations down to pretty low $\pt \simeq 4$~GeV values in order to compare with data, we shall take slightly larger scales, $\muoptpi = \pt / \sqrt{2}$ and $\muoptga = \pt / 2$, than the strict requirement Eq.~(\ref{eq:opt}). It appears moreover that using $\muoptga = \pt / 2$ allows for an excellent description of prompt-photon measurements, from fixed-target to collider experiments~\cite{Fontannaz:2006pc}. The theoretical uncertainties discussed in the following Section are estimated from the scale-variation around these central values.

\subsection{Proton--proton collisions}
\label{sec:pp}

\begin{figure}[ht]
  \begin{minipage}[ht]{7.8cm}
    \begin{center}
      \includegraphics[height=7.8cm]{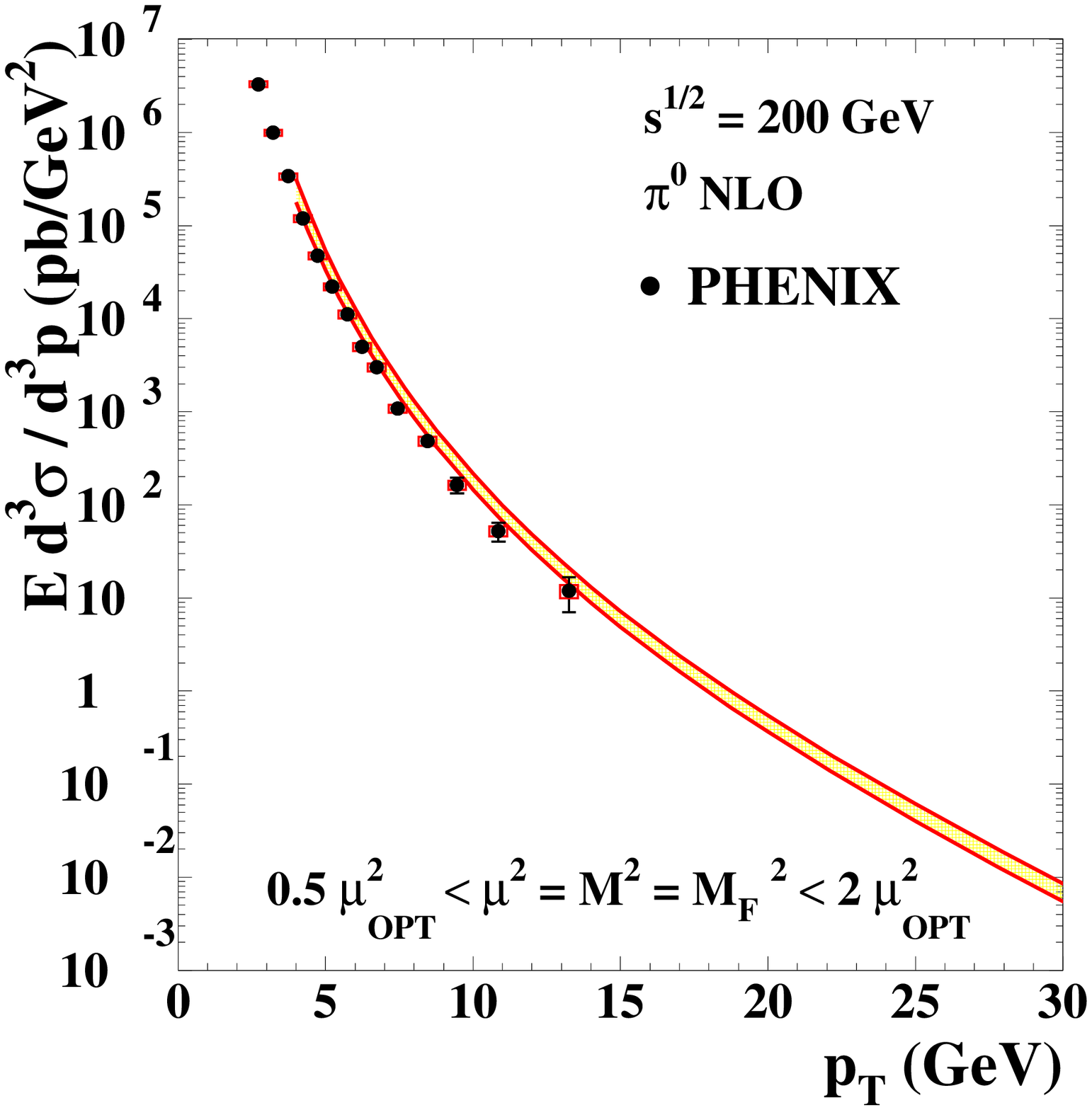}
    \end{center}
  \end{minipage}
~
  \begin{minipage}[ht]{7.8cm}
    \begin{center}
      \includegraphics[height=7.8cm]{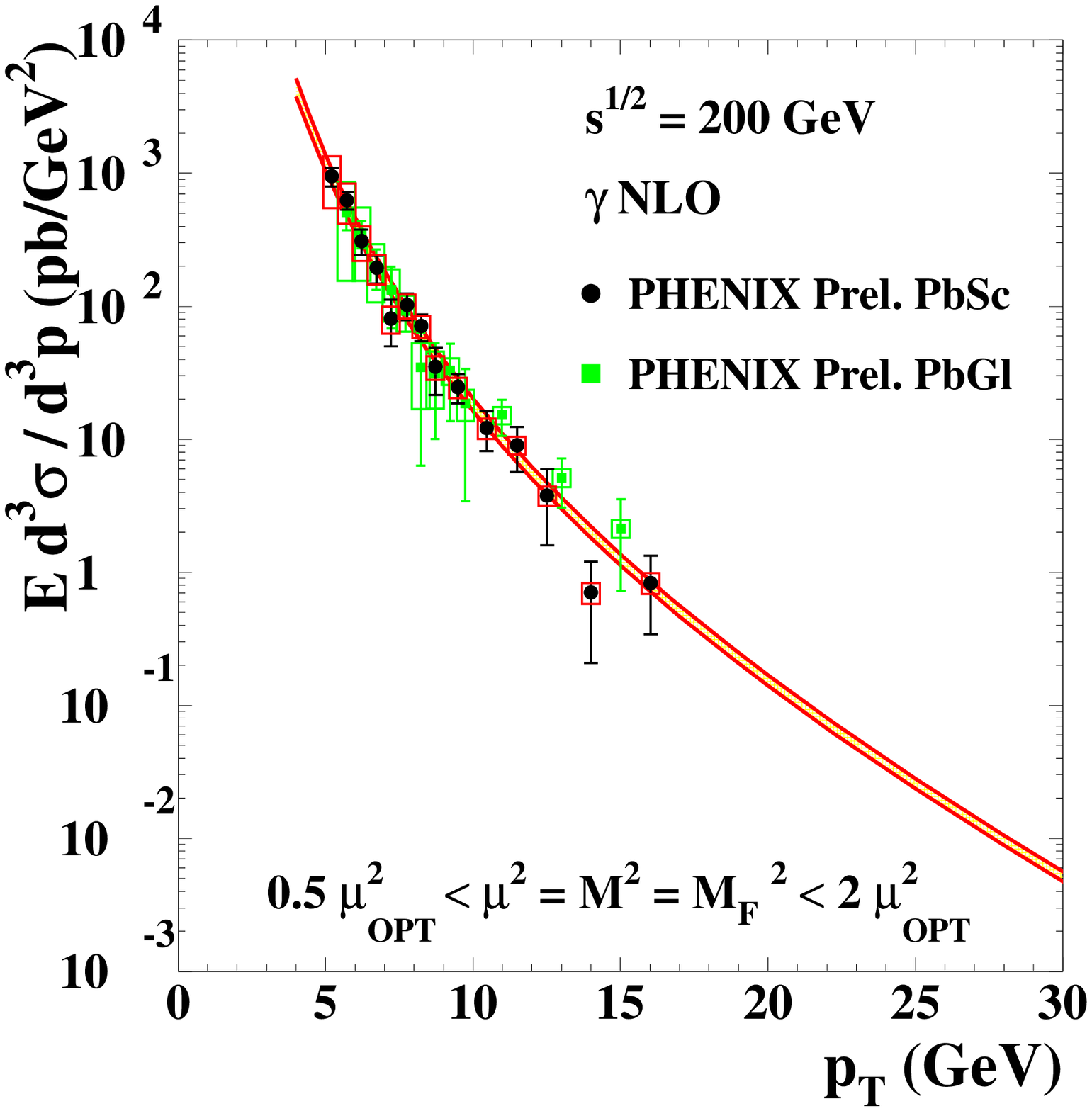}
    \end{center}
  \end{minipage}
  \caption{Single-pion (left) and single-photon (right) invariant cross section at mid-rapidity in $p$--$p$ collisions computed at NLO accuracy, varying simultaneously the factorization, the renormalization and the fragmentation scales from $\muopt/\sqrt{2}$ to $\sqrt{2} \, \muopt$. The PHENIX data for pions~\cite{Adler:2003pb} (9.6\% normalization error not shown) and the PHENIX preliminary data for photons~\cite{Okada:2005in} are also shown for comparison. Photons and pions are produced in the $[-0.35;0.35]$ rapidity interval.}
  \label{fig:spectra}
\end{figure}

The single-inclusive NLO production in $p$--$p$ collisions is presented in this Section. In Fig.~\ref{fig:spectra} are shown the pion (left) and the prompt photon (right) $p_{_\perp}$ spectra. Predictions are given from $p_{_\perp} \ge$~4 GeV --~below which perturbative calculations are not expected to be reliable~-- and up to $p_{_\perp} =$ 30 GeV, above which cross sections are too small to be measured with the current RHIC luminosity. At small $p_{_\perp}$, pion production cross sections prove almost two orders of magnitude, that is $\cO{\alpha_s/\alpha}$, larger than prompt photon cross sections. At larger $p_{_\perp}\gtrsim 20$~GeV, however, both processes turn out to have a similar yield, since the fragmentation mechanism in the pion channel becomes kinematically disfavoured with respect to the direct photon process.

\begin{figure}[ht]
  \begin{minipage}[ht]{7.8cm}
    \begin{center}
      \includegraphics[height=7.8cm]{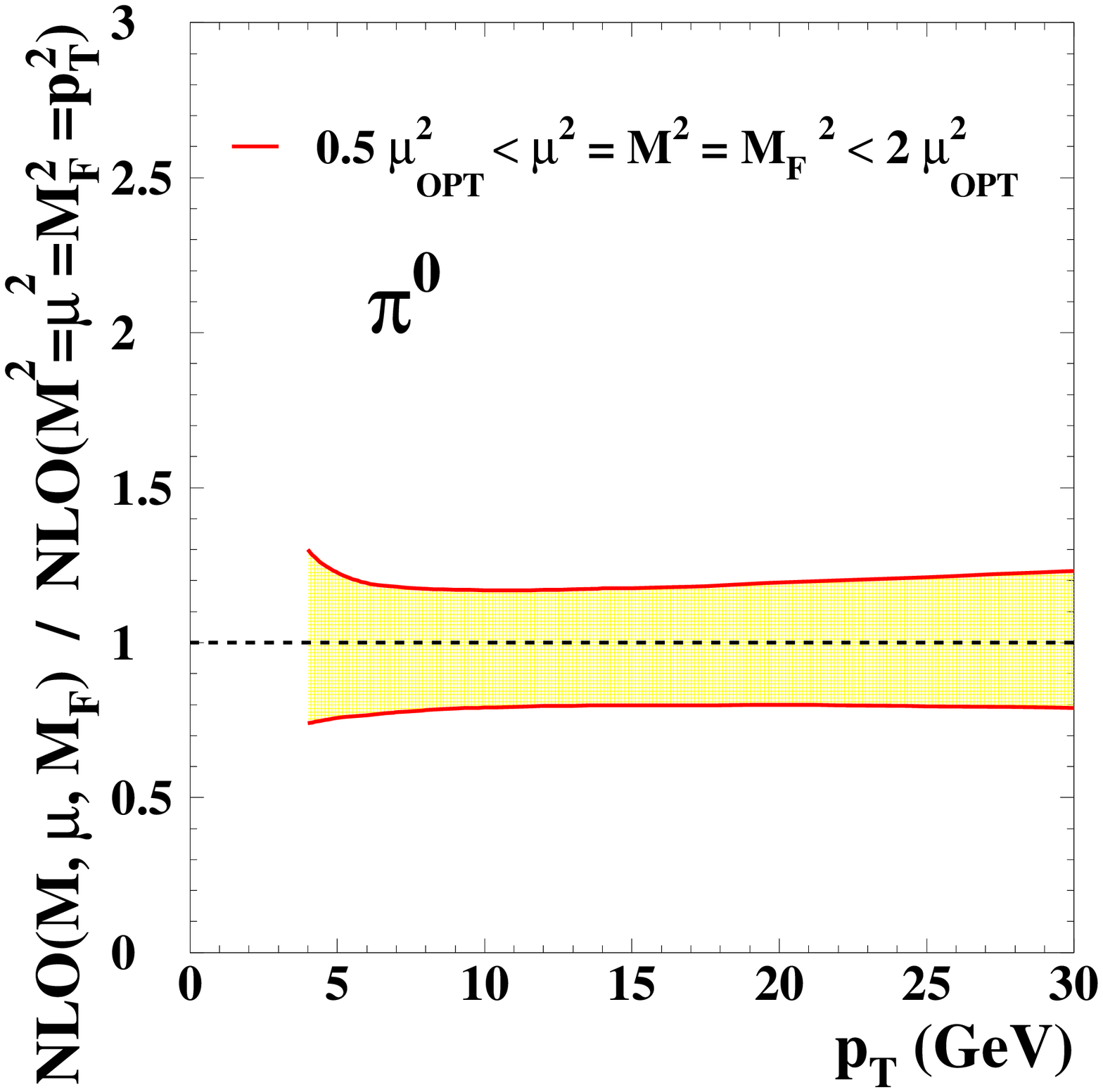}
    \end{center}
  \end{minipage}
~
  \begin{minipage}[ht]{7.8cm}
    \begin{center}
      \includegraphics[height=7.8cm]{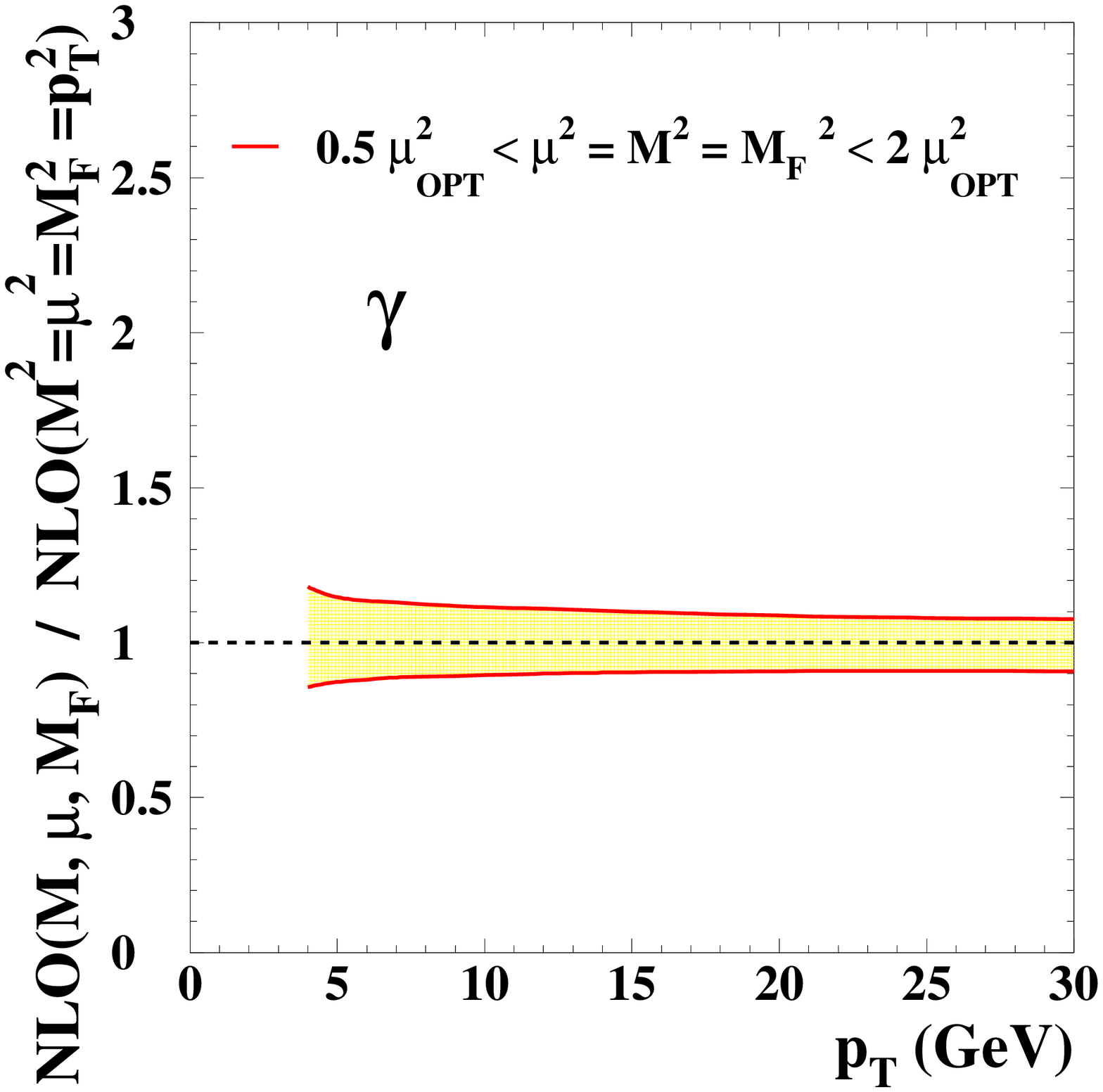}
    \end{center}
  \end{minipage}
  \caption{Scale dependence of the NLO predictions for single-pion (left) and single-photon (right) production at RHIC. Predictions varying simultaneously all scales from $\muopt/\sqrt{2}$ to $\sqrt{2} \muopt$ are normalized to the ``central'' prediction $\mu = M = M_{_F} = \muopt$.}
  \label{fig:scale}
\end{figure}
\begin{figure}[ht]
  \begin{minipage}[ht]{7.8cm}
    \begin{center}
    \includegraphics[height=7.8cm]{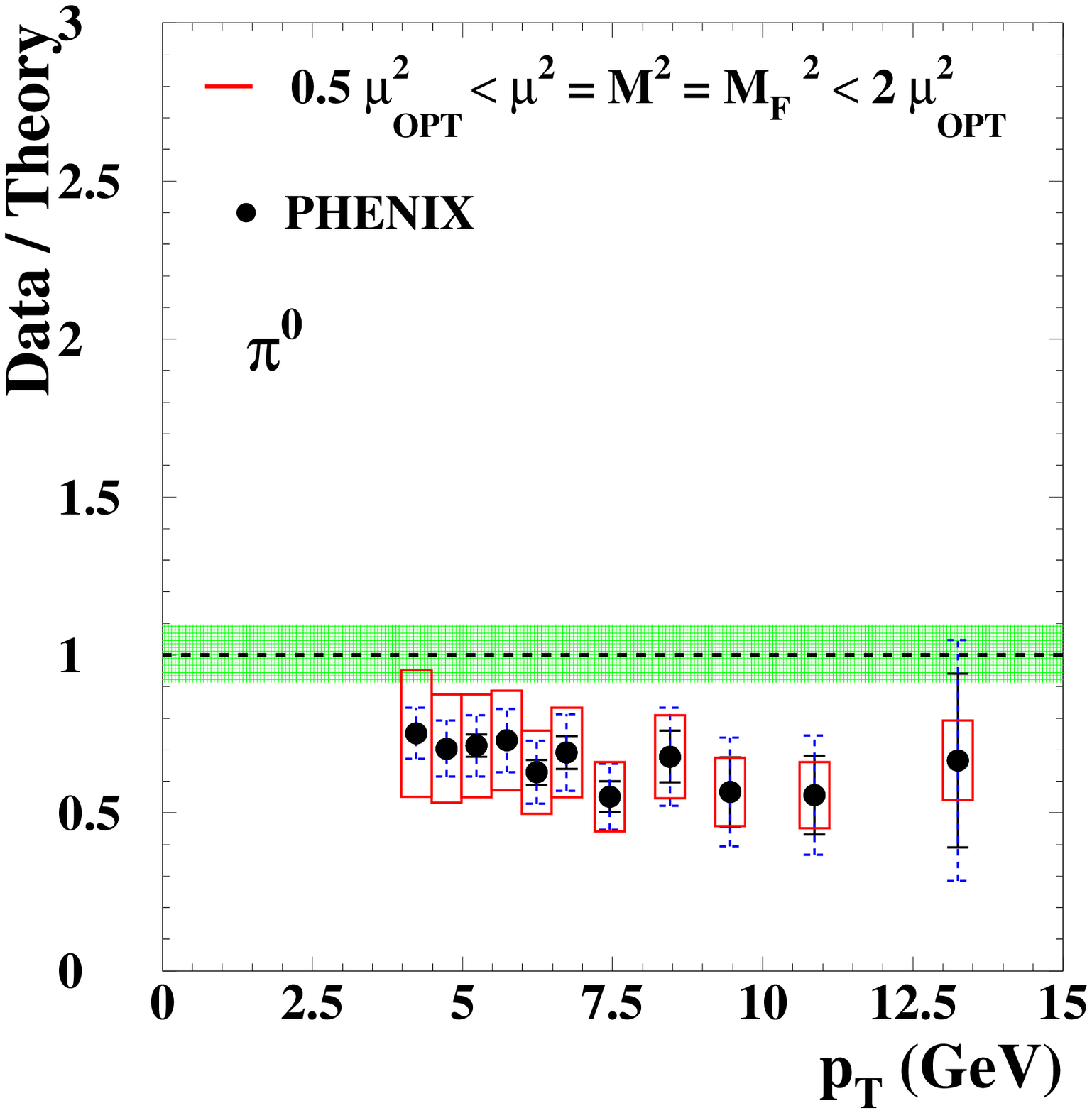}  
    \end{center}
  \end{minipage}
~
  \begin{minipage}[ht]{7.8cm}
    \begin{center}
      \includegraphics[height=7.8cm]{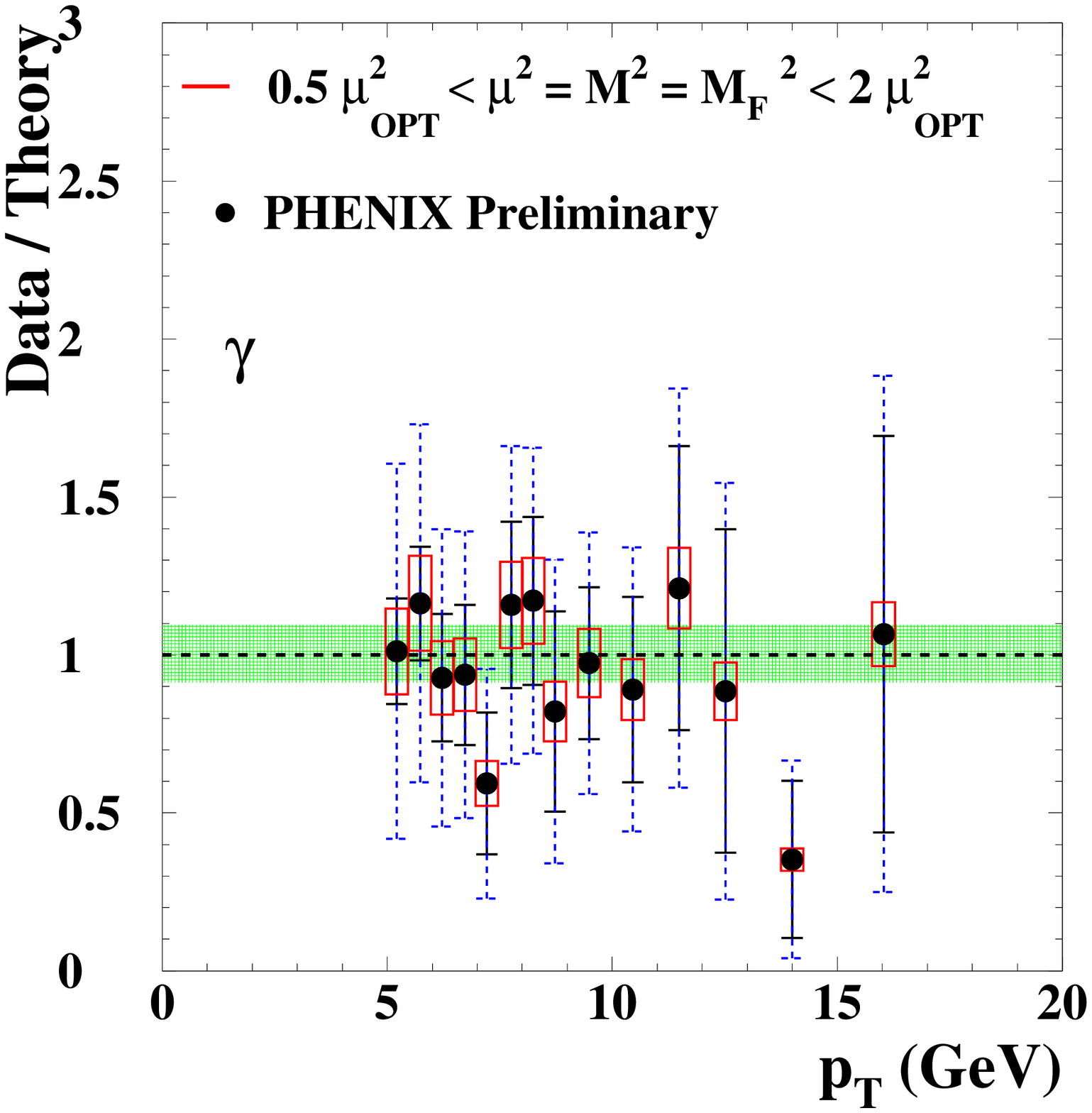}
    \end{center}
  \end{minipage}
  \caption{PHENIX data for single-pion (left) and PHENIX preliminary single-photon data (right) normalized to the NLO predictions. The experimental error bars are shown as solid (statistical) and dashed (statistical plus systematic) lines, and the theoretical systematic uncertainty is shown as a box. The band indicates the 9.6\% normalization error in the PHENIX measurements.}
  \label{fig:dataovth}
\end{figure}

The band in Fig.~\ref{fig:spectra} represents the theoretical systematic error of the NLO calculation coming from the scale-fixing procedure. In order to discuss this uncertainty on a more quantitative level, Fig.~\ref{fig:scale} displays the ratio of the NLO cross sections at scales varying from $\muopt /\sqrt{2}$ to $\sqrt{2} \, \muopt$, normalized to the ``central'' prediction, all scales being equal to $\muopt$ (see Sect.~\ref{sec:scalefixing}). This uncertainty turns out to be rather stable as a function of $\pt$ for both processes, although the magnitude in both channels differs somewhat. For pions, varying the scales affects cross sections by roughly 30\%, while a smaller 15\% dependence is observed in prompt photon production. The reason is twofold: on the one hand, pion production requires an additional $\alpha_s$ coupling (hence a larger renormalization scale dependence); on the other hand, the direct photon process does not show any dependence on the fragmentation scale. Note that varying scales independently (e.g. crossing the scales $\mu = \sqrt{2}\, \muopt$ and $M_{_F} = \muopt /\sqrt{2}$) leads to predictions that lie within our uncertainty band\footnote{P. Aurenche and M. Werlen, private communication.}. We should repeat, however, that the photon fragmentation functions are much less constrained than those in the pion sector; this leads to an additional theoretical uncertainty, not quantified here.

Let us now discuss the pion and photon NLO predictions in comparison with the recent PHENIX measurements~\cite{Adler:2003pb,Okada:2005in,Frantz:2004gg}. As can already be seen in Fig.~\ref{fig:spectra}, an excellent agreement is found on the whole $\pt$ range. To be more precise, the data over the NLO theory ratio is computed in Fig.~\ref{fig:dataovth}. Boxes indicate the theoretical uncertainty on this ratio coming from the scale dependence, and the vertical error bars show the statistical uncertainty in PHENIX data, with (dashed) or without (solid) systematic error. 

For pions, it is worth noting that errors coming from theory and experiment remain not too large over the whole $\ptpi$ range. At small $\ptpi$, the main uncertainty is given by the scale dependence of the NLO computation (approximately 30\%) while experimental error bars prove as small as 10\%. On the contrary, at large $\ptpi$, the error on the data-over-theory ratio is mostly statistical, because of the low counting rates with the present RHIC luminosity. Fig.~\ref{fig:dataovth} clearly indicates that the present NLO $\pi^0$ cross sections overestimate the PHENIX measurements by roughly 30\%; yet the data-over-theory ratio proves to be flat over the whole $\ptpi$ range. Slightly larger scales, say all equal to $\ptpi$, would significantly improve the description of PHENIX data. It is nevertheless quite remarkable to observe such a fair agreement between the NLO predictions and the data since the KKP fragmentation functions used here have not been constrained by any hadron--hadron scattering data~\cite{Kniehl:2000fe}. It may therefore certainly be useful to reverse the logics and to use the neutral-pion PHENIX data in a global (N)LO QCD fit analysis to further constrain the pion fragmentation functions and thus gain additional accuracy on the LHC predictions. 

Let us move to prompt photon production (Fig.~\ref{fig:dataovth}, right). The data-over-theory ratio is now completely dominated by the statistical uncertainty in the PHENIX data at all transverse momenta. Theoretical predictions prove to be in excellent agreement with data\footnote{On the contrary, the $\mu=M=M_{_F}=\ptgamma$ prescription shown in~\cite{Okada:2005in} somehow underestimates the PHENIX measurements.}. We hope that the experimental errors can be reduced in the future, so as to test more drastically the current NLO prompt photon predictions, whose fragmentation functions are not well constrained yet.

\subsection{Nucleus--nucleus collisions}
\label{sec:aasingle}

\begin{figure}[ht]
  \begin{minipage}[ht]{7.8cm}
    \begin{center}
      \includegraphics[height=7.8cm]{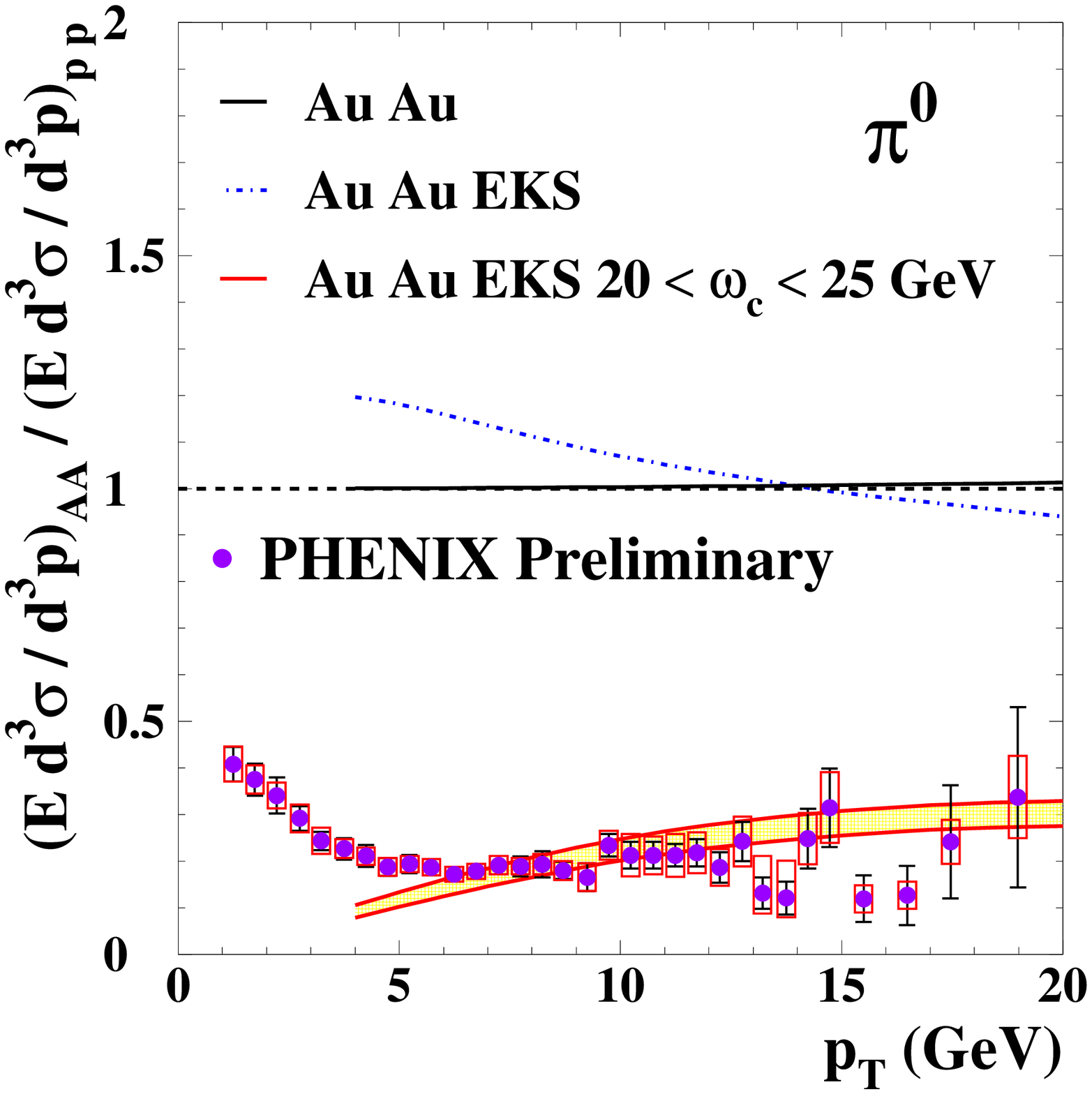}  
    \end{center}
  \end{minipage}
~
  \begin{minipage}[ht]{7.8cm}
    \begin{center}
      \includegraphics[height=7.8cm]{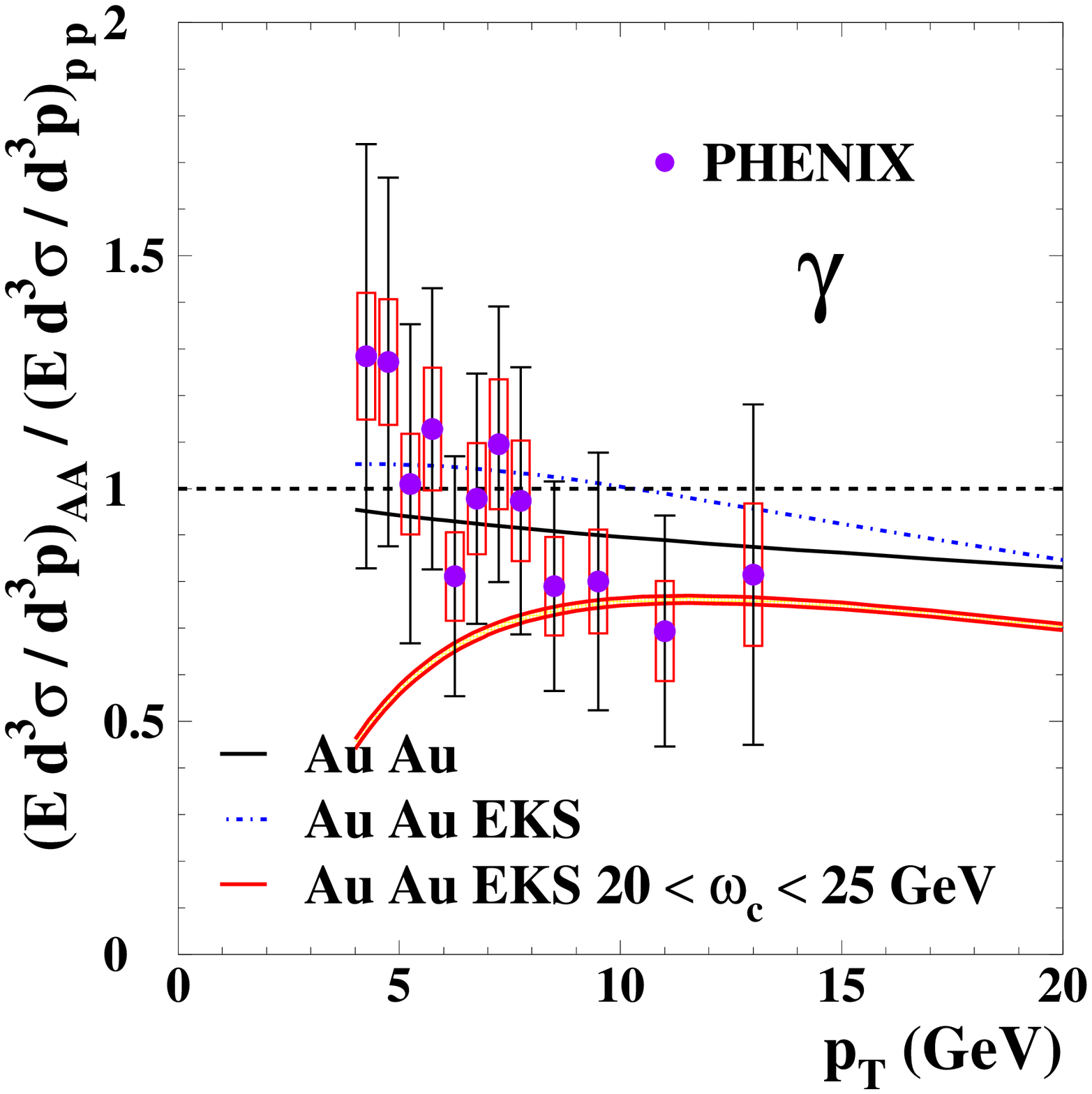}  
    \end{center}
  \end{minipage}
  \caption{Ratio of Au--Au over $p$--$p$ single-pion (left) and single-photon (right) production cross section. Calculations are done at LO, assuming (i) isospin (solid), (ii) isospin and shadowing (dotted), (iii) isospin, shadowing and energy loss (band) effects (see text for details). The PHENIX preliminary data on the single-pion~\cite{Isobe:2005pc} and single-photon~\cite{Adler:2005ig} production in Au--Au collisions normalized respectively to $p$--$p$ scattering $\pi^0$ data and to the present NLO prompt photon prediction in $p$--$p$ collisions are also shown for comparison.}
  \label{fig:quenchingsingle}
\end{figure}

We now come to the production of single-$\pi^0$ and single-$\gamma$ production in Au--Au collisions. In Fig.~\ref{fig:quenchingsingle} is plotted the expected quenching of these two probes (respectively on the left and on the right panel) assuming (i) no nuclear effect to be at work (labelled ``Au~Au'', solid), (ii) including shadowing corrections (``Au~Au EKS'', dash-dotted), and (iii) including both shadowing and energy loss processes (``Au~Au EKS $20 < \omega_c < 25$ GeV'', band). 

While isospin effects in the pion channel are negligible at RHIC, we note that antishadowing tends to enhance the nuclear production by roughly 20\% in the lowest $\ptpi$  bin. Conversely, at large $\ptpi \gtrsim 15$~GeV, the large-$x$ EMC effect slightly quenches the pion yield. The single-pion spectrum in Au--Au collisions including shadowing is therefore somewhat softened with respect to the $p$--$p$ scattering case, although the net effect remains small. In contrast, the effect of the parton energy loss process, shown as a band, is dramatic. The quenching factor starts around 0.1 at $\ptpi =~4$~GeV and smoothly increases to 0.3 at $\ptpi =~20$~GeV, in rather good agreement with the PHENIX preliminary data~\cite{Isobe:2005pc}. Note that, even if the $\ptpi$ dependence of the $\pi^0$ quenching is somehow flattened by the shadowing contribution, no hint for such an increase is seen in the data. The rather flat behaviour observed there may be due to the geometrical bias introduced when integrating over all possible path lengths~\cite{Eskola:2004cr}, not performed here.

In contradistinction to the pion case, the isospin correction is significant in prompt photon production (right, solid). Photon production being an electromagnetic process, cross sections depend on the light-quark electric charges and are thus disfavoured in a nucleus target, less rich in up quarks than a proton is. It is worth stressing that the quenching -- without any shadowing nor energy loss effects -- proves as large as 20\% at $\ptgamma = 20$~GeV!  Antishadowing slightly compensates the isospin effect to make the quenching factor closer to unity below 10 GeV. Unlike the pion channel, antishadowing extends up to $\ptgamma \simeq 20$ GeV, at which it is negligible. This is so because the Bjorken $x$ probed in the {\it direct} photon channel is smaller, $x = \cO{2 \ptgamma / \sqrt{s}}$, than in the pion fragmentation process, $x = \cO{2 \ptpi / z_{_\pi} \sqrt{s}}$. 

Let us now comment on the energy loss effects. Because of the dominance of the direct process unaffected by the medium (keeping in mind that its relative contribution is scale-dependent and thus somehow arbitrary),  the photon quenching is not as pronounced as that of the pion. At 4 GeV, prompt photon production is suppressed by 40\% but remains of order 30\% in the largest considered $\ptgamma$ bins. While energy loss effects in the present model lead to an increase of the quenching factor with $\ptgamma$, the rather flat behavior seen in Fig.~\ref{fig:quenchingsingle} actually comes from the interplay of the energy loss process on the one hand and of the isospin effects on the other hand.

The quenching of single-photon production is unfortunately not yet available due to the too low statistics in $p$--$p$ collisions. In Fig.~\ref{fig:quenchingsingle} (right) is shown the ratio of the Au--Au PHENIX photon measurements~\cite{Adler:2005ig} to the present NLO calculations in $p$--$p$ collisions. In that respect, this is not --~strictly speaking~-- the {\it same} quenching factor as the one determined in the pion sector for instance. Nevertheless, given the good agreement between PHENIX data and NLO calculations discussed in Section~\ref{sec:pp} (cf. Fig.~\ref{fig:dataovth}, right), we believe this ratio to be indicative of the genuine prompt photon quenching (i.e. normalized to $p$--$p$ {\it data}). Interestingly, we notice that the photon quenching factor turns out to be in very good agreement with the PHENIX measurements. Although the presently too large error bars do not allow one to disentangle the predictions with or without energy loss effects, it is worth stressing that our estimated 30\% suppression is not inconsistent with these preliminary measurements. It would indeed be particularly intriguing not to observe a suppression --~though less spectacular than for the pions~-- due to the parton energy loss. Calculations done at RHIC using isolation criteria~\cite{Okada:2005in} indeed indicate that roughly 20\% of the inclusive photon yield actually comes with an important hadronic activity\footnote{I thank P. Aurenche and S. Bathe for discussions on this issue.} (``jet-like photons''). Let us mention that our predictions follow the same trend as the calculation by Jeon, Jalilian-Marian, and Sarcevic~\cite{Jeon:2002dv} who first attempted to determine prompt photon quenching at RHIC. However, we can regret the lack of clear relationship between the energy loss probability distribution used in~\cite{Jeon:2002dv} and the medium-induced gluon radiation computed in QCD by BDMPS~\cite{Baier:1997sk}.

The difference between the single-pion and the single-photon quenching discussed above can be better seen in the ratio of the photon total yield --~i.e. including prompt photons as well as the background photons coming from the decay of neutral pions~-- over that very background:
\be
R_{\gamma / \pi^0} = \frac{\sigma(p p \to \gamma \,\X) + \sigma(p p \to \pi^0 \,\X \to \gamma \,\X)}{\sigma(p p \to \pi^0 \,\X \to \gamma \,\X)}.
\ee
The $\pi^0$ decay contribution $\pi^0 \to \gamma \,\X$ to the photon total yield may be simply estimated from the slope $n$ of the single-pion spectra assuming a power law behaviour\footnote{The slope $n$, and therefore the ratio $R_{_{{\rm decay}}}$, is estimated  from the PHENIX preliminary data in $p$--$p$ collisions~\cite{Adler:2003pb}. Since the quenching factor is remarkably flat above 4~GeV, $n$ should not change much from $p$--$p$ to Au--Au collisions.} $d\sigma/dp_{_{\perp}} \sim p_{_\perp}^{-n}$,~\cite{Arleo:2003gn,Ferbel:1984ef}
\be
\frac{\sigma (p p \to \pi^0 \,\X \to \gamma \,\X)}{\sigma (p p \to \pi^0 \,\X)} = \frac{2}{n-1}.
\ee

\begin{figure}[ht]
  \centering
  \includegraphics[height=7.8cm]{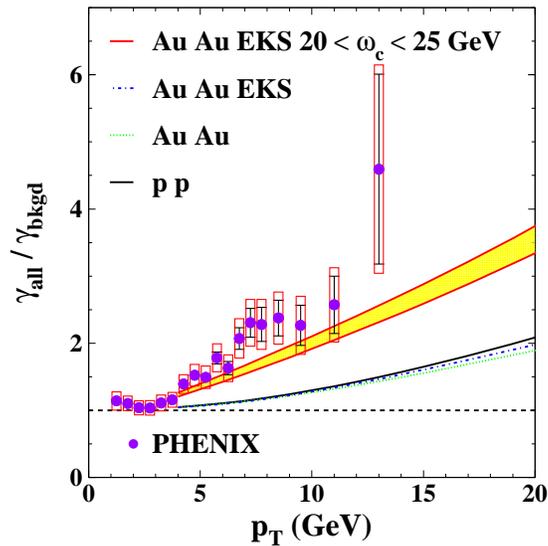}
  \caption{Ratio of the overall photon production over the background photon production coming from the $\pi^0$ decay. LO calculations are done in $p$--$p$ and Au--Au collisions (see text for details). The PHENIX data~\cite{Adler:2005ig} are shown for comparison.}
  \label{fig:gammaoverpion}
\end{figure}

In $p$--$p$ collisions (Fig.~\ref{fig:gammaoverpion}, solid line), the ratio increases from 1 at small transverse momentum --~that is with a negligible prompt photon production~-- up to 2 at 20 GeV, above which the prompt photon signal takes over the background. Shown respectively as a dash-dotted and as a dotted line are the same ratios in Au--Au collisions when adding isospin effect, respectively with and without shadowing corrections. The overall effect of these cold nuclear matter corrections is pretty weak, making the ratio slightly smaller than what is expected in $p$--$p$ collisions.

On the contrary, one might expect dramatic effects of parton energy loss on this observable. Indeed, since the pion yield proves more suppressed by this process than does the photon yield, this ratio should increase much faster with $\ptgamma$, in the presence of the hot medium. The ratio has been computed assuming the previous $\omega_c=20$--$25$~GeV range in Fig.~\ref{fig:gammaoverpion} (band). As expected, $R_{\gamma / \pi^0}$ turns out to be almost twice as large at 20 GeV in Au--Au collisions with energy loss than in $p$--$p$ collisions. Moreover, it is particularly interesting to notice that this prediction proves to be in good agreement with the recent PHENIX data in central Au--Au collisions, although the predicted slope tends to underestimate that of the data. Nevertheless, we find it quite remarkable that both single inclusive $\pi^0$ and $\gamma$ channels can be described within the same model and with one common parameter. This allows us to use such a model further and to consider, in Section~\ref{sec:double}, photon-tagged correlations as a possible tool to probe more quantitatively the medium produced at RHIC. 

\subsection{Discussion}
\label{sec:discussion}

\subsubsection{RHIC energy density}
\label{sec:rhicdensity}

The good agreement between the single-inclusive $p$--$p$ and Au--Au scattering data and the theoretical calculations is a first basic check of the present approach. Perhaps more importantly, this allows us to constrain the unknown parameter
 of the model, $\omega_c$, and then to relate it to other physical quantities regarding the dense-medium produced. We would like here to briefly comment on its absolute value in order to get a rough estimate of the energy density currently achieved in central Au--Au collisions at RHIC.

The dynamical scaling law sets the relation between the time-averaged $\langle \hat{q} \rangle$ and the initial-time $\hat{q}(t_0)$ transport coefficient~\cite{Salgado:2002cd}:
\begin{eqnarray}
  \label{eq:scalinglaw}
  \langle \hat{q} \rangle &=& \frac{2}{L^2}\, \int_{t_0}^L \, dt \,(t-t_0)\,\left(\frac{t_0}{t}\right)^\alpha\,\hat{q}(t_0),  \nonumber \\
& \simeq & \frac{2}{2-\alpha} \,\left(\frac{t_0}{L}\right)^\alpha\,\hat{q}(t_0) \qquad {\rm when}\,\,\, t_0 \ll L,
\end{eqnarray}
accounting for the expansion of the produced medium with density $n(t) \propto t^{-\alpha}$. Assuming in the following a purely longitudinal expansion ($\alpha = 1$) in (\ref{eq:scalinglaw}), the transport coefficient at an initial time $t_0$ is thus given by~\cite{Baier:2002tc}
\begin{equation}
  \label{eq:omctoq}
  \hat{q}(t_0) \simeq \frac{\omega_c}{t_0\,L}.
\end{equation}
Using the estimate $\omega_c = 20$--$25$~GeV in~(\ref{eq:omctoq}), one then gets the transport coefficient $\hat{q}(t_0 = 0.5~\rm{fm}) \simeq 1.6$--$2$~GeV$^2$/fm at an early time $t_0 = 0.5$~fm after the reaction.

\begin{figure}[ht]
  \centering
  \includegraphics[height=9.cm]{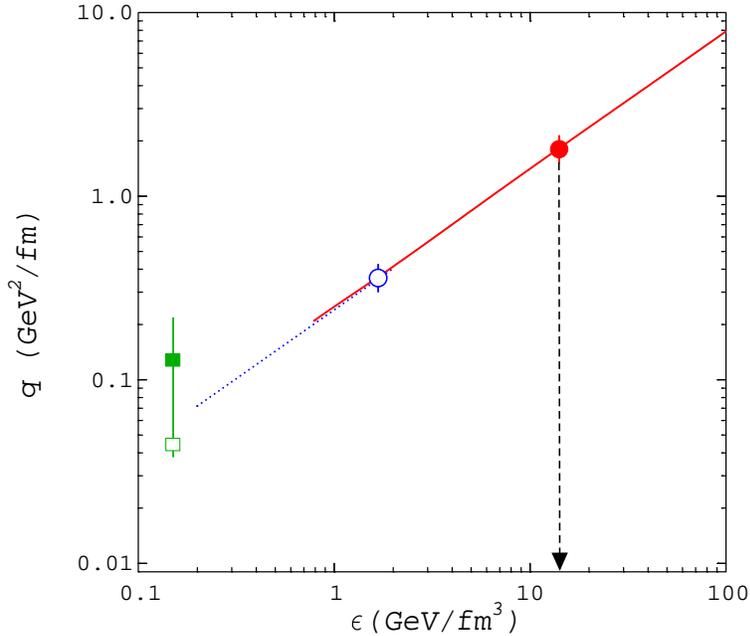}
  \caption{The solid (respectively, dashed) line represents the energy density dependence of the transport coefficient in a thermally equilibrated quark--gluon plasma (respectively, pion gas)~\cite{Baier:2002tc}. The squares indicate the transport coefficient for cold nuclear matter computed perturbatively (open square)~\cite{Baier:1997sk} and extracted from Drell-Yan and DIS data (full square)~\cite{Arleo:2003jz,Arleo:2002ph}. The open and full circles represent respectively the mean $\langle \hat{q} \rangle$ and the initial-time $\hat{q}$($t_0=0.5$~fm) transport coefficient at RHIC (figure adapted from~\cite{Baier:2002tc}, with permission).}
  \label{fig:qhat}
\end{figure}

In Ref.~\cite{Baier:2002tc}, Baier related the transport coefficient computed perturbatively~\cite{Baier:1997sk} to the medium energy density, $\epsilon$, both for hot pion gas or quark--gluon plasma. Fig.~\ref{fig:qhat} exhibits the generic power law dependence of the transport coefficient on $\epsilon$. In order to make a crude estimate on the energy density reached in central Au--Au collisions at RHIC energy, we superimposed on this curve our estimates for the time-averaged (open circle) and initial time (full circle) transport coefficient. This leads to an energy density of about $\epsilon \simeq 15$~GeV/fm$^3$, i.e. 100 times that of cold nuclear matter. For comparison, also shown on this plot is the nuclear matter transport coefficient, both the perturbative estimate (open square)~\cite{Baier:1997sk} as well as the slightly larger but consistent value extracted from Drell--Yan production in pion-induced nuclear collisions~(full square)~\cite{Arleo:2002ph}.

Let us recall here that our estimate is no more than a guess, as it may somehow depend on our specific assumptions, regarding for instance the longitudinal expansion, the choice to take a fixed medium length, not to say the intrinsic uncertainty coming from the perturbative calculation of the transport coefficient. Nevertheless, we notice that our result, $\epsilon_{_{\rm{RHIC}}} / \epsilon_{_{\rm{cold}}} \simeq 100$, proves to be on the same order of magnitude as estimates based on soft particle production at mid-rapidity~\cite{Back:2004je}. Moreover, this number should be seen as a lower estimate, since our assumptions --~in particular the geometrical modelling as well as the absence of a vanishing cut-off in the BDMPS medium-induced gluon spectrum~-- is likely to underestimate the initial-time transport coefficient $\hat{q}(t_0)$. 

It would also be interesting to get an estimate on both temperature and energy density from the above transport coefficient within a quasi-particle picture. The gluon damping rate $\gamma$, or equivalently its inverse mean free path ${\lambda_g}^{-1}$, has been computed in SU($N_c$) colour gauge theory in the weak coupling limit~\cite{Pisarski:1993rf}. It is given by
\begin{equation}\label{eq:damping}
  \gamma = {\lambda_g}^{-1} = \frac{1}{2} \, \alfas \, N_c \, T \, \ln\left(\frac{c}{\alfas}\right).
\end{equation}
Fitting the entropy computed on the lattice in quenched QCD, Peshier extracted the coefficient appearing in the logarithm of~(\ref{eq:damping}), $c =2.23$~\cite{Peshier:2004bv}. From the leading-order Debye mass $\mD = \sqrt{4 \,\pi \alfas}\, T$, we obtain the relation between transport coefficient and temperature:
\begin{equation*}
  \hat{q} = \mD^2 \, \gamma = 6 \, \pi \, \alfasd \, T^3 \, \ln\left( \frac{c}{\alfas} \right).
\end{equation*}
Taking the transport coefficient $\hat{q} = 1.6$--$2$~GeV$^2$/fm and $\alfas = 1/2$ at rather soft scales, the temperature of the medium is $T = 355$--$385$~MeV. In a gas of weakly interacting gluons, this would correspond to an energy density $\epsilon_{_{\rm RHIC}} = 16 \,\pi^2 \, T^4 / 30 = 11$--$15$~GeV/fm$^3$, which compares well with the former estimate.


\subsubsection{Uncertainties on medium-induced photon production}
\label{sec:photonfragmentation}

Despite the good agreement between the model predictions and the PHENIX data, the currently large experimental and theoretical uncertainties do not rule out other possible channels for photon production in a hot medium. 

To leading order, the photon can participate directly in the hard partonic subprocess and thus escape without any strong interaction. On top of that direct channel, photon production may also come from the collinear fragmentation of quarks and gluons produced in the hard process. Although of higher order according to power counting rules, the large logarithm $\ln (Q^2/\Lambda^2) \sim \alpha_s^{-1}$ coming from the integral over the photon transverse momentum makes this process contribute to the same leading order $\cO{\alpha \, \alfas}$ in the perturbative expansion as the direct process~\cite{Aurenche:1998gv}. The formation time needed to produce such a photon--parton system with a small invariant mass exceeds by far the typical lifetime of the hot medium. This allowed us to consider a two-stage process: the multiple scattering of the hard parton in the medium is followed on a much larger time-scale by the parton-to-photon fragmentation process in the vacuum. 

However, the possibility for the photon to be produced while still inside the medium has not been considered here. In particular, the multiple scattering incurred by hard partons induces the emission of soft gluons {\it as well as} soft photons. Within his path-integral picture for the parton energy loss mechanism~\cite{Zakharov:1996fv,Zakharov:1997uu}, Zakharov recently computed this medium-induced photon bremsstrahlung contribution at RHIC~\cite{Zakharov:2004bi}. The enhancement of photon production is particularly noticeable in the moderate $\ptgamma$ range, say below 20~GeV, and could therefore somewhat balance the photon quenching predicted in Fig.~\ref{fig:quenchingsingle}. In addition, large transverse momentum partons may couple to the thermal quarks and gluons in the medium through Compton scattering or $q\bar{q}$ annihilation. This jet--photon conversion mechanism, considered by Fries, M\"uller and Srivastava in Ref.~\cite{Fries:2002kt}, is particularly rich since it would directy reflect the hard {\it parton} instead of the measured hadron spectrum, provided this channel is dominant in photon production.

Several analyses came out, recently, which embed some --~or all~-- of these channels into hydrodynamical evolution~\cite{Turbide:2005fk,d'Enterria:2005vz,Huovinen:2001wx,Gelis:2004ep}. The agreement with the recent PHENIX data~\cite{Okada:2005in} is remarkable down to low $\pt$ values, $\pt \gtrsim 1$~GeV. In particular, it appears that the new preliminary and more precise data~\cite{Bathe:2005pc} are better reproduced with a thermal contribution~\cite{d'Enterria:2005vz}. This is clearly an important observation, which would deserve further study. Note, in passing, that a strong depletion of the perturbative QCD yield expected at low $\ptgamma$ because of the energy loss mechanism would make the photon enhancement even much stronger! However, one should not expect the NLO pQCD calculations to be reliable and well constrained at such a small $\pt$. In order to bypass the comparison with the pQCD calculation extrapolation, it would be extremely interesting to measure the quenching factor down to 1~GeV and to see whether the photon enhancement remains.

Each of these individual mechanisms is unfortunately poorly under control, not to mention the uncertainty due to the space-time evolution of the expanding medium. Given this variety of processes, it appears that the medium effects on (prompt) photon production still remain quite unknown and model-dependent. Hopefully more precise data in the near future will shed light on the prompt photon production mechanism in nucleus-nucleus collisions. In the following, however, we shall discuss double inclusive pion--photon correlations at some specific regions of phase space in which the photon is (mostly) produced directly in the hard process. Consequently, our predictions should fortunately not depend much on the medium-induced photon production process.

\section{Double inclusive photon-pion production}
\label{sec:double}

The momentum imbalance
\begin{equation*}
 \z = - \frac{{\bf p_{_{\perp_{\gamma}}}} \ .\ {\bf p_{_{\perp_{\pi}}}}} {p^2_{_{\perp_{\gamma}}}}
\end{equation*}
spectrum between a hard prompt photon and a much softer (but still hard) hadron produced in hadronic collisions may allow for the determination of the hadron fragmentation function, $D_i^h(z\simeq z_{_{3 4}})$, the photon transverse momentum balancing that of the parton $i$, which fragments into the hadron. At least this two-body kinematics may be a valid picture when higher-order corrections, briefly discussed below, remain small. Moreover, the fixed-order calculations should not be reliable at large $\z$ because of the large logarithms $\alfas^n \ln^{2n}(1-\z)$ and $\alfas^n \ln^{2n-1}(1-\z)$, due to soft and collinear gluon emissions, which need to be resummed to all orders.

This triggered a recent phenomenological study of various pion--photon and photon--photon correlations in heavy-ion collisions at LHC energy~\cite{Arleo:2004xj}. Our goal here is to investigate similarly such momentum correlations at RHIC energy, restricting ourselves to the $\gampi$ channel. For more details concerning the other kinematical variables used here, see~\cite{Arleo:2004xj}.

\subsection{Proton--proton collisions}

In order to probe the pion fragmentation function efficiently through the $\gampi$ momentum imbalance spectrum, asymmetric cuts $\picut \ll \gacut$ are required to make the range covered in the $z$ fragmentation variable as wide as possible. In addition, the pion momentum needs to be hard enough to ensure the perturbative regime to be at work, while the photon momentum should not be too large to maintain reasonable counting rates:
\be
\Lambda_{_{\rm QCD}} \ll \picut \ll \gacut \ll \sqrt{s} \,/2.
\ee
In the following we choose the cuts $\picut=3$~GeV and $\gacut=10$~GeV.

\begin{figure}[ht]
  \centering
   \includegraphics[height=14.cm]{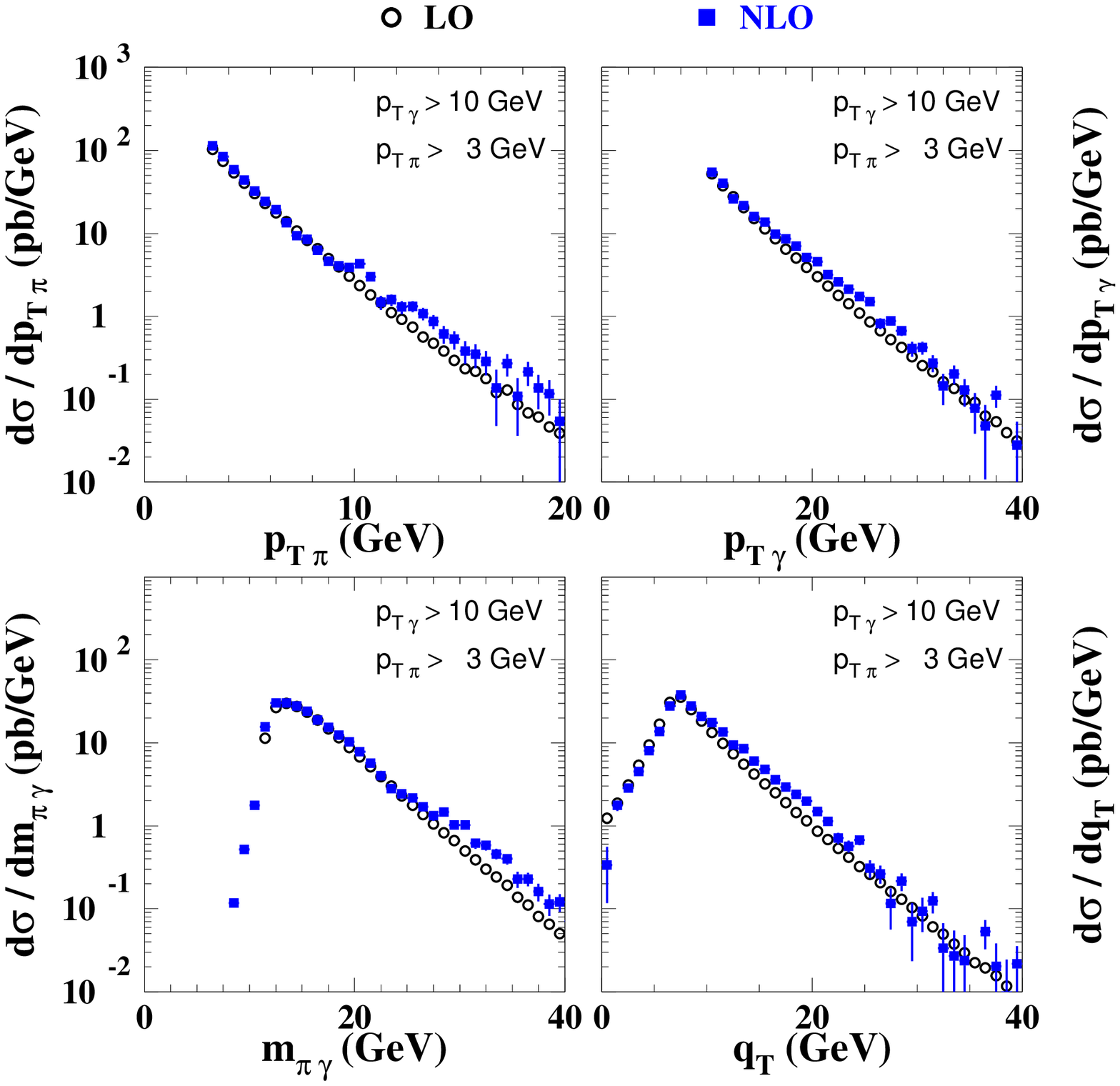}
  \caption{The pion $\ptpi$ and the photon $\ptgamma$ transverse momentum distributions (top), the invariant mass $\m$, and the pair transverse momentum $\qt$ (bottom) are computed in proton--proton scattering to LO (open circles) and NLO (full squares) at $\sqrt{s} = 200$~GeV. Both the photon and the pion are produced at rapidity $[-0.5;0.5]$ and the cuts $\gacut~=~10~$GeV and $\picut =3$~GeV are imposed.}
\label{fig:quarteron_pp}
\end{figure}

Fig.~\ref{fig:quarteron_pp} shows the pion and the photon $p_{_\perp}$ spectra (top) as well as the invariant mass $\m$ and the transverse momentum of the pair $\qt$ (bottom) computed in $p$--$p$ collisions at LO (open circles) and NLO (full squares) accuracy. To leading order, the two particles are emitted back to back in the parton--parton center-of-mass frame; hence the invariant mass distribution has a threshold given by $2 \sqrt{\picut \, \gacut} \simeq 11$~GeV and the $\qt$ spectrum shows a maximum around $\gacut - \picut = 7$~GeV. At NLO, new configurations in momentum space show up since the pion and the photon may be emitted at a smaller relative azimuthal angle (we apply a minimal cut $\phigampi \ge \pi /2$). This fills in particular the small invariant mass region (with a new threshold $\sqrt{\picut \, \gacut} \simeq 5.5$~GeV) and shifts the $q_{_T}$ spectrum to slightly larger values. Except in this region of phase space, however, LO predictions appear quite stable with respect to the small NLO corrections, say roughly 20--30\% at most. We should however keep in mind that the strength of NLO corrections strongly depends on the renormalization, factorization and fragmentation scales assumed in the pQCD calculation (here, all fixed at $\ptgamma/2$).

\begin{figure}[ht]
  \centering
   \includegraphics[height=9.cm]{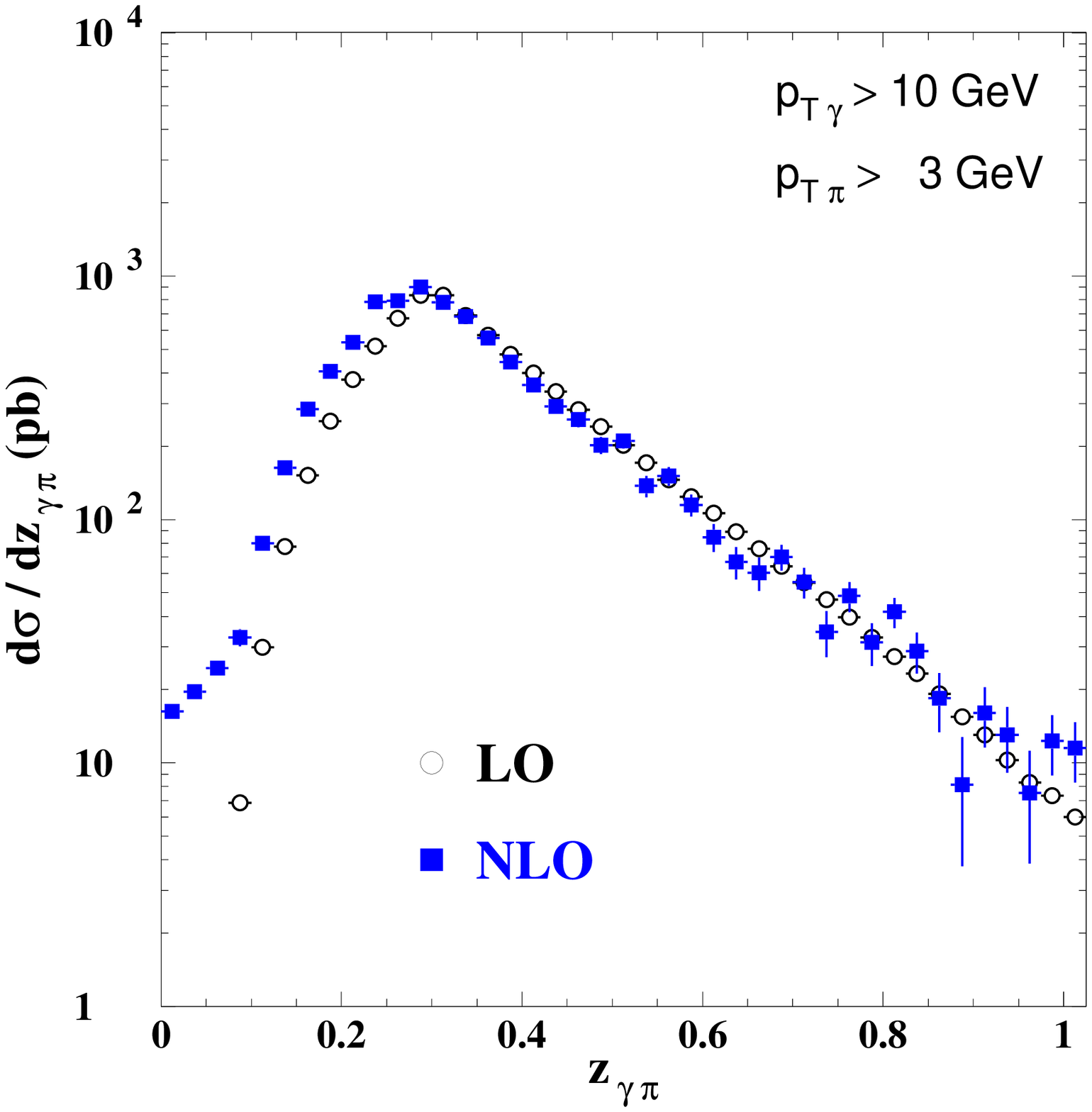}
  \caption{The $\gampi$ momentum imbalance $\z$ distribution is computed in proton--proton scattering to LO (open circles) and NLO (full squares) at $\sqrt{s} = 200$~GeV. Both the photon and the pion are produced at rapidity $[-0.5, 0.5]$ and the cuts $\gacut = 10~$GeV and $\picut =3$~GeV are imposed.}
\label{fig:z_pp}
\end{figure}

The distribution in the $\gampi$ momentum imbalance $\z$ is also determined to LO and NLO accuracy (Fig.~\ref{fig:z_pp}). As already pointed out in Ref.~\cite{Arleo:2004xj}, its shape turns out to be reminiscent of the pion fragmentation function. Above the ratio of the cuts, $\picut / \gacut = 0.3$, the distribution decreases as the pion momentum and thus the fragmentation variable $z$ gets larger. Similarly, the distribution is strongly suppressed below $\z < 0.3$, since larger photon momenta are needed, keeping $\ptpi$ close to its lower cut. Again, higher-order corrections prove large when the two particles are emitted at $\phigampi \gtrsim \pi/2$, thus at small values of $\z$  (recall that $\z \propto \cos \phigampi$), but moderate elsewhere. Let us furthermore insist that, despite the regular behaviour of the NLO predictions, soft gluon resummation may affect the distribution at high $\z$.

\subsection{Nucleus--nucleus collisions}
\label{sec:aadoublespectra}

\begin{figure}[ht]
  \centering
  \includegraphics[height=14.cm]{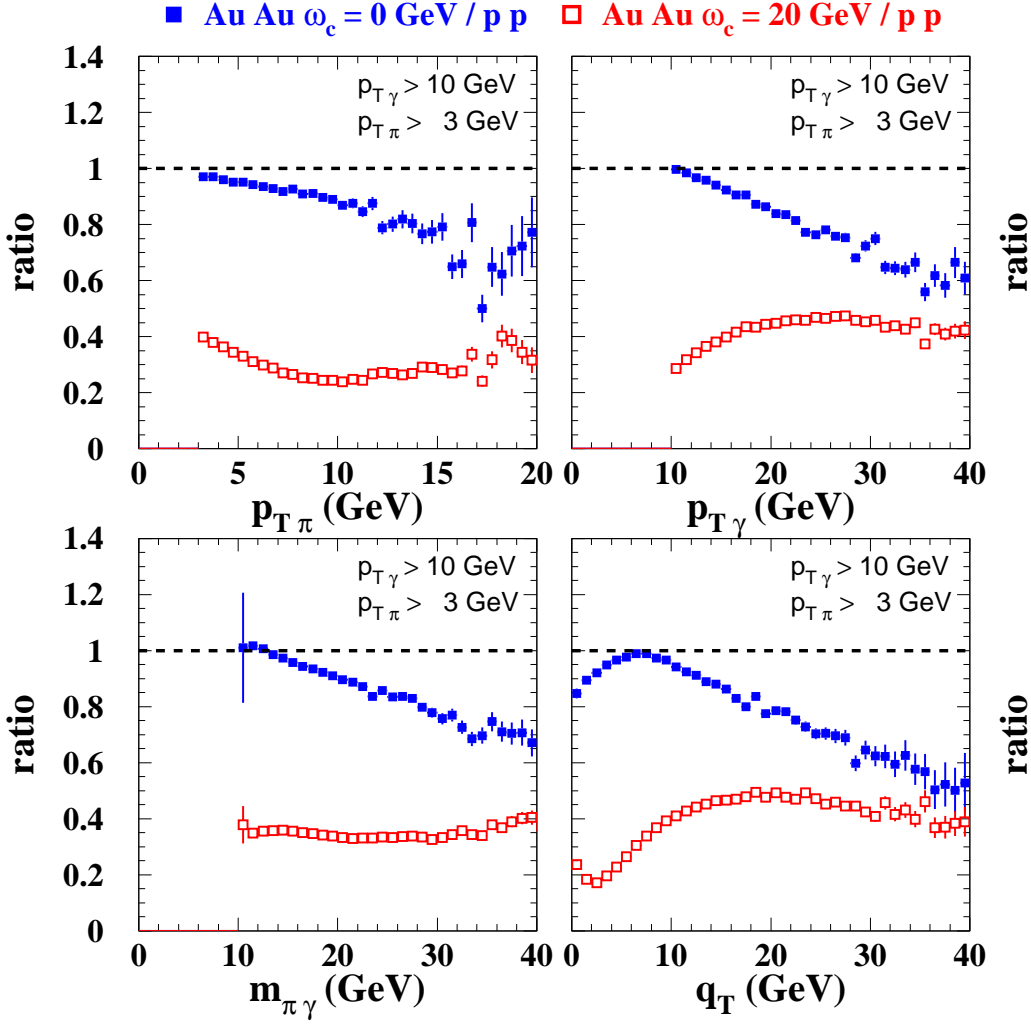}
  \caption{Same distributions as Fig.~8. Distributions are computed in Au--Au collisions at $\sqrt{s} = 200$~GeV assuming shadowing (full squares) and shadowing with energy loss  (open squares) normalized to the $p$--$p$ distributions.}
  \label{fig:quarteron_quench}
\end{figure}

In Fig.~\ref{fig:quarteron_quench} the quenching factors of the $\ptpi$, $\ptgamma$, $\qt$ and $\m$ LO spectra in Au--Au collisions are computed, including nuclear shadowing and with energy loss ($\omega_c = 20$~GeV, open squares) or without it ($\omega_c = 0$~GeV, full squares). The effect of isospin and nuclear shadowing is rather small for all observables when both the pion $\ptpi$ and the photon $\ptgamma$ momenta are close to their respective cuts, i.e. where distributions are maximal. However, the quenching factor decreases down to 0.7 when momenta become larger (e.g. at $\ptgamma \simeq 40$~GeV) since higher $x$ are probed in the Au nuclei. Just as the single production case, the quenching comes from the interplay between the lack of up quarks in the nuclei and the nuclear EMC effect. Note that the depletion is also pronounced at small $\qt$, when the pion momentum is of the order of the photon momentum, $\ptpi \simeq \ptgamma$. In this specific region, not only the pion but also the photon are produced by collinear fragmentation. This double fragmentation process then requires highly energetic (i.e. large-$x$) partons to be produced to fragment into the two detected particles, therefore leading to a similar suppression.

\

The effect of the energy loss on each of the spectra is more dramatic. In particular, unlike isospin and shadowing, the smaller the momenta the stronger the effects. As an example, the quenching is as large as 0.3 close to the photon momentum cut, $\ptgamma \gtrsim 10$~GeV, or even 0.2 at small pair momentum, $\qt \simeq 5$~GeV. As in the single production case, energy loss effects naturally die out at larger momenta, since the relative energy loss, $\epsilon / \kt$, gets smaller. Once more, though, the competition between energy loss on the one hand and isospin/shadowing on the other leads to various patterns for the quenching. While the invariant mass quenching remains remarkably flat on the whole kinematic range, the quenching slightly decreases (respectively, increases) with $\ptpi$ (respectively, $\ptgamma$). Particularly interesting is the $\qt$ spectrum behaviour in the medium. As stressed above, increasing $\qt$ above its maximum (at $\picut - \gacut$) amounts to increasing the photon momentum. This is the reason why both the $\ptgamma$ and the $\qt$ spectra are similarly quenched (Fig.~\ref{fig:quarteron_quench}, left). At very small and negative $\qt$, however, the pion becomes harder while the photon momentum remains of order $\gacut$. Since the double fragmentation process makes energy loss effects weaker, the quenching turns out quite naturally to be maximal around $\gacut-\picut$. 

\begin{figure}[ht]
  \centering
  \includegraphics[height=9.cm]{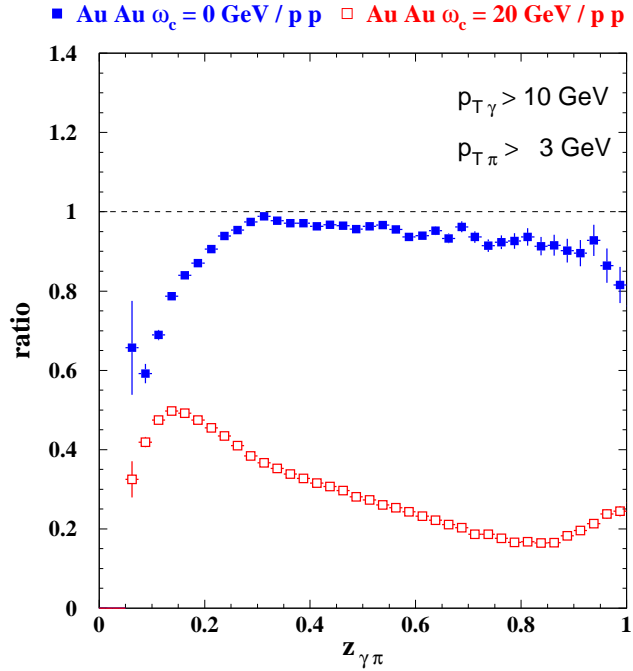}
  \caption{The $\gampi$ momentum imbalance $\z$ distributions is computed in Au--Au collisions at $\sqrt{s} = 200$~GeV assuming shadowing (full squares) and shadowing with energy loss  (open squares) normalized to the $p$--$p$ distributions.}
  \label{fig:z_quench}
\end{figure}

Let us finally discuss the $\gampi$ momentum imbalance quenching shown in Fig.~\ref{fig:z_quench}. Remarkably, the effect of isospin/shadowing is completely negligible (much less than 10\%) above the cut ratio, $\z \gtrsim \picut/\gacut$. In that domain, the pion momentum $\ptpi$ grows while the photon momentum --~and therefore the transverse momentum of the parton that fragments, to leading order~-- is kept fixed. The typical values of $x$ do not change much and hence neither does the quenching due to isospin or shadowing. These effects are then located in the small region $\z \lesssim \picut/\gacut$ where the photon momentum is much larger, $\ptgamma\gg\gacut$. The energy loss mechanism leads to a completely different picture for the quenching. At larger $\z \simeq z$, the phase space for gluon emission is dramatically restricted (see Eq.~(\ref{eq:modelFF})) and the quenching becomes more pronounced. This explains the decreasing behaviour of the quenching factor, from 0.5 at $\z \simeq 0.1$ down to 0.15 at $\z\simeq 0.8$. At very large $\z > 0.8$, the quenching factor starts increasing again because of the onset of the double fragmentation process, as already mentioned. 

It is clearly the momentum imbalance variable that offers the largest observable difference whether including energy loss ($\omega_c = 20$~GeV) or not ($\omega_c = 0$~GeV) in the model. Fig.~\ref{fig:z_quench} also justifies a posteriori our choice of extremely asymmetric cuts between the pion and the photon momenta, in order to isolate as much as possible the effects of isospin and nuclear shadowing.

\

\subsection{Counting rates}
\label{sec:rates}

The momentum correlations between a (ideally directly produced) prompt photon and a pion, as discussed above, are clearly of interest at RHIC energy in order to probe and constrain the present energy loss models, which have described successfully the single pion/photon production. Hopefully double inclusive $\gampi$ data can be available in the near future at RHIC. 

For this reason, we would like to discuss here what the expected counting rates are for such processes. The number of events is given by~\cite{Arleo:2003gn}
\begin{equation*}
{\cal N}^{{\rm hard}}_{_{{\rm Au Au}}}\big |_{_{\cal C}} = {\cal L}_{_{\rm int}} \,\times\, \langle N_{{\rm coll}} \rangle \big |_{_{\cal C}} \,\, \frac{\sigma^{{\rm geo}}_{_{{\rm AuAu}}}}{\sigma_{_{{\rm NN}}}} \,\, \sigma^{{\rm hard}}_{_{{\rm NN}}} \,\,\, {\cal C},
\label{eq:nucl-sec}
\end{equation*}
The RHIC highest integrated luminosity delivered in Run~4 and measured by PHENIX is taken to be ${\cal L} = 0.7~{\rm nb}^{-1}$~\cite{Ahrens:2005pc}. Cross sections in Au--Au collisions range roughly from $10^{-4}$ to $10^{-1}$~nb/GeV, so that approximately ${\cal N} = 2$-$2~10^3$~events/GeV are expected. Regarding the imbalance distribution, rates should be $\dd {\cal N}/\dd z = 5~10^2$-$5~10^4$ events. These numbers are given obviously without considering any additional kinematic cuts or acceptance restrictions. A higher luminosity at RHIC is clearly hoped for, as it would certainly allow for a more systematic investigation of $\gampi$ or $\gamgam$ correlations with a great  variety of cuts.

\section{Summary}

Single inclusive production of pions and photons at large transverse momentum ($p_{_\perp}\ge~4$~GeV) has been computed perturbatively in $p$--$p$ and Au--Au collisions at RHIC energy ($\sqrt{s}=200$~GeV). In $p$--$p$ collisions, NLO calculations compare successfully with the PHENIX data on the whole $p_{_\perp}$ range for both pion and photon spectra. Theoretical uncertainty from the scale fixing in the calculation was also examined. In Au--Au collisions, we determined the $p_{_\perp}$ spectra, assuming possible shadowing corrections in the nuclear parton densities or energy loss effects on the fragmentation process. The quenching of single-pion production observed by the PHENIX collaboration can be reproduced, assuming the typical scale $\omega_c = 20$--$25$~GeV for the energy loss process. From these values, we estimate the energy density reached at RHIC in central Au--Au collisions to be $\epsilon_{_{\rm RHIC}} \gtrsim 10$~GeV/fm$^3$ at an initial (and somewhat arbitrary) time $t_0 = 0.5$~fm. Within the same energy loss model, the quenching factor proves much less pronounced in the photon sector, because of the presence of the direct process channel unaffected by the medium. The expected photon quenching is found to be roughly 20\%, therefore in agreement with the PHENIX measurements. Finally, the ratio of the photon total yield over the pion decay background was found to fairly reproduce the present data.

The production of double inclusive $\gampi$ momentum correlations has then been investigated in detail. In $p$--$p$ scattering, LO predictions were shown to be quite stable with respect to higher-order corrections, except in some specific regions of phase space. It is the case in particular when pions and photons are emitted at small relative azimuthal angle, thus at small momentum imbalance. Using the asymmetric cuts $\picut = 3$~GeV and $\gacut = 10$~GeV, the quenching of various correlation spectra in Au--Au collisions are predicted. The momentum imbalance distribution $\z$ is seen to be particularly sensitive to the medium-modified fragmentation dynamics. The expected counting rates assuming the largest RHIC integrated luminosity in Run-4 are encouraging, even though a higher luminosity would be required for a thorough study of photon-tagged correlations at RHIC. 

As an outlook, it would be interesting to go beyond the present calculation and to perform a systematic comparison of photon-tagged versus hadron-tagged momentum and azimuthal correlations. Since the latter is shown to be quite sensitive to the surface emission, and hence to the geometry of the collision~\cite{Eskola:2004cr}, one could naturally expect significant differences in such correlations when triggering on prompt photons. Going from a coloured (parton or fragmentation photon) to a blind (direct photon) trigger would scan different energy density profiles. In that respect, performing isolated versus non-isolated photon correlation measurements would be ideal.

\acknowledgments

I would like to thank warmly Patrick Aurenche for many discussions and suggestions on this work as well as for his hospitality at LAPTH. I am also indebted to Stefan Bathe, Jean-Philippe Guillet, and Dominique Schiff for useful comments and discussion. David d'Enterria is also particularly acknowledged for his careful reading of the manuscript.

\providecommand{\href}[2]{#2}\begingroup\raggedright
\endgroup


\begin{thebibliography}{10}

\bibitem{Feinberg:1976ua}
E.~L. Feinberg, {\it Direct production of photons and dileptons in
  thermodynamical models of multiple hadron production},  Nuovo Cim. {\bf A34}
  (1976) 391.

\bibitem{Shuryak:1978ij}
E.~V. Shuryak, {\it Quark--gluon plasma and hadronic production of leptons,
  photons and psions},  Phys. Lett. {\bf B78} (1978) 150.

\bibitem{Aggarwal:2000th}
{\bf WA98} Collaboration, M.~M. Aggarwal {\em et~al.}, {\it Observation of
  direct photons in central 158 {AGeV} {Pb} + {Pb} collisions},  Phys. Rev.
  Lett. {\bf 85} (2000) 3595--3599
  [\href{http://arXiv.org/abs/nucl-ex/0006008}{{\tt nucl-ex/0006008}}].

\bibitem{Arleo:2003gn}
F.~Arleo {\em et~al.}, {\it Photon physics in heavy ion collisions at the
  {LHC}},  Report of the ``Photon Physics Working Group'' in the CERN Yellow
  Report on ``Hard Probes in Heavy Ion Collisions at the LHC'', CERN--2004--009
  [\href{http://arXiv.org/abs/hep-ph/0311131}{{\tt hep-ph/0311131}}].

\bibitem{Akiba:2005pc} {\bf {PHENIX}} Collaboration, Y.~Akiba, {\it
Probing the properties of dense partonic matter at RHIC}, talk given
at {Quark Matter} 2005, {Budapest} ({Hungary}), 4-9 {August} 2005,
\href{http://arXiv.org/abs/nucl-ex/0510008}{{\tt nucl-ex/0510008}}.

\bibitem{Bathe:2005pc} {\bf {PHENIX}} Collaboration, S.~Bathe, {\it
Direct photon in 200 {GeV} p+p, d+{Au}, {Au}+{Au} from PHENIX}, talk
given at {Quark Matter} 2005, {Budapest} ({Hungary}), 4-9 {August}
2005, \href{http://arXiv.org/abs/nucl-ex/0511042}{{\tt
nucl-ex/0511042}}.

\bibitem{Owens:1986mp}
J.~F. Owens, {\it Large momentum transfer production of direct photons, jets,
  and particles},  Rev. Mod. Phys. {\bf 59} (1987) 465.

\bibitem{Aurenche:1987fs}
P.~Aurenche, R.~Baier, M.~Fontannaz and D.~Schiff, {\it Prompt photon
  production at large {$p_{_\perp}$} scheme invariant {QCD} predictions and
  comparison with experiment},  Nucl. Phys. {\bf B297} (1988) 661.

\bibitem{Aurenche:1992yc}
P.~Aurenche, P.~Chiappetta, M.~Fontannaz, J.~P. Guillet and E.~Pilon, {\it
  Next-to-leading order bremsstrahlung contribution to prompt photon
  production},  Nucl. Phys. {\bf B399} (1993) 34--62.

\bibitem{Gordon:1993qc}
L.~E. Gordon and W.~Vogelsang, {\it Polarized and unpolarized prompt photon
  production beyond the leading order},  Phys. Rev. {\bf D48} (1993)
  3136--3159.

\bibitem{Aurenche:1998gv}
P.~Aurenche {\em et~al.}, {\it A critical phenomenological study of inclusive
  photon production in hadronic collisions},  Eur. Phys. J. {\bf C9} (1999)
  107--119 [\href{http://arXiv.org/abs/hep-ph/9811382}{{\tt hep-ph/9811382}}].

\bibitem{Binoth:1999qq}
T.~Binoth, J.-P. Guillet, E.~Pilon and M.~Werlen, {\it A full next-to-leading
  order study of direct photon pair production in hadronic collisions},  Eur.
  Phys. J. {\bf C16} (2000) 311--330
  [\href{http://arXiv.org/abs/hep-ph/9911340}{{\tt hep-ph/9911340}}].

\bibitem{Aurenche:1999nz}
P.~Aurenche, M.~Fontannaz, J.-P. Guillet, B.~A. Kniehl and M.~Werlen, {\it
  Large {$p_{_\perp}$} inclusive {$\pi^0$} cross-sections and next-to-leading
  order {QCD} predictions},  Eur. Phys. J. {\bf C13} (2000) 347--355
  [\href{http://arXiv.org/abs/hep-ph/9910252}{{\tt hep-ph/9910252}}].

\bibitem{Iancu:2003xm}
E.~Iancu and R.~Venugopalan, {\it The {Color Glass Condensate} and high energy
  scattering in {QCD}},  {Quark-Gluon Plasma 3, R.C. Hwa and X.-N. Wang Ed.,
  World Scientific} (2003) 240--361
  [\href{http://arXiv.org/abs/hep-ph/0303204}{{\tt hep-ph/0303204}}].

\bibitem{Baier:2000mf}
R.~Baier, D.~Schiff and B.~G. Zakharov, {\it Energy loss in perturbative
  {QCD}},  Ann. Rev. Nucl. Part. Sci. {\bf 50} (2000) 37--69
  [\href{http://arXiv.org/abs/hep-ph/0002198}{{\tt hep-ph/0002198}}].

\bibitem{Gyulassy:2003mc} M.~Gyulassy, I.~Vitev, X.-N.~Wang,
B.-W.~Zhang, {\it Jet quenching and radiative energy loss in dense
nuclear matter}, {Quark-Gluon Plasma 3, R.C. Hwa and X.-N. Wang Ed.,
World Scientific} (2003) 123--191
[\href{http://arXiv.org/abs/nucl-th/0302077}{{\tt nucl-th/0302077}}].

\bibitem{Adcox:2001jp}
{\bf {PHENIX}} Collaboration, K.~Adcox {\em et~al.}, {\it Suppression of
  hadrons with large transverse momentum in central {Au} + {Au} collisions at
  {$\sqrt{s}$} = 130 {GeV}},  Phys. Rev. Lett. {\bf 88} (2002) 022301
  [\href{http://arXiv.org/abs/nucl-ex/0109003}{{\tt nucl-ex/0109003}}].

\bibitem{Adler:2002xw}
{\bf STAR} Collaboration, C.~Adler {\em et~al.}, {\it Centrality dependence of
  high {$p_{_\perp}$} hadron suppression in {Au} + {Au} collisions at
  {$\sqrt{s}$} = 130 {GeV}},  Phys. Rev. Lett. {\bf 89} (2002) 202301
  [\href{http://arXiv.org/abs/nucl-ex/0206011}{{\tt nucl-ex/0206011}}].

\bibitem{Adler:2003qi}
{\bf PHENIX} Collaboration, S.~S. Adler {\em et~al.}, {\it Suppressed {$\pi^0$}
  production at large transverse momentum in central {Au} + {Au} collisions at
  {$\sqrt{s}$} = 200 {GeV}},  Phys. Rev. Lett. {\bf 91} (2003) 072301
  [\href{http://arXiv.org/abs/nucl-ex/0304022}{{\tt nucl-ex/0304022}}].

\bibitem{Adler:2002tq}
{\bf STAR} Collaboration, C.~Adler {\em et~al.}, {\it Disappearance of
  back-to-back high {$p_{_\perp}$} hadron correlations in central {Au} + {Au}
  collisions at {$\sqrt{s}$} = 200 {GeV}},  Phys. Rev. Lett. {\bf 90} (2003)
  082302 [\href{http://arXiv.org/abs/nucl-ex/0210033}{{\tt nucl-ex/0210033}}].

\bibitem{Adams:2004wz}
{\bf STAR} Collaboration, J.~Adams {\em et~al.}, {\it Azimuthal anisotropy and
  correlations at large transverse momenta in p + p and {Au} + {Au} collisions
  at {$\sqrt{s}$} = 200 {GeV}},  Phys. Rev. Lett. {\bf 93} (2004) 252301
  [\href{http://arXiv.org/abs/nucl-ex/0407007}{{\tt nucl-ex/0407007}}].

\bibitem{Dietel:2005st} {\bf STAR} Collaboration, T.~Dietel, {\it
Azimuthal correlations of high {$p_{_\perp}$} photons and hadrons in
{Au} + {Au} collisions at {RHIC}}, talk given at Quark Matter 2005,
Budapest (Hungary), 4-9 Aug 2005,
\href{http://arXiv.org/abs/nucl-ex/0510046}{{\tt nucl-ex/0510046}}.

\bibitem{Chiu:2002ma}
{\bf PHENIX} Collaboration, M.~Chiu, {\it Charged particle angular correlations
  from leading photons at {RHIC}},  Nucl. Phys. {\bf A715} (2003) 761--764
  [\href{http://arXiv.org/abs/nucl-ex/0211008}{{\tt nucl-ex/0211008}}].

\bibitem{Adler:2005ee}
{\bf PHENIX} Collaboration, S.~S. Adler {\em et~al.}, {\it Modifications to
  di-jet hadron pair correlations in {Au} + {Au} collisions at {$\sqrt{s}$} =
  200 {GeV}},  \href{http://arXiv.org/abs/nucl-ex/0507004}{{\tt
  nucl-ex/0507004}}.

\bibitem{Wang:2003mm}
X.-N. Wang, {\it High {$p_{_\perp}$} hadron spectra, azimuthal anisotropy and
  back-to-back correlations in high-energy heavy-ion collisions},  Phys. Lett.
  {\bf B595} (2004) 165--170 [\href{http://arXiv.org/abs/nucl-th/0305010}{{\tt
  nucl-th/0305010}}].

\bibitem{Qiu:2003pm}
J.~Qiu and I.~Vitev, {\it Transverse momentum diffusion and broadening of the
  back- to-back di-hadron correlation function},  Phys. Lett. {\bf B570} (2003)
  161--170 [\href{http://arXiv.org/abs/nucl-th/0306039}{{\tt
  nucl-th/0306039}}].

\bibitem{Wang:1996yh}
X.-N. Wang, Z.~Huang and I.~Sarcevic, {\it Jet quenching in the opposite
  direction of a tagged photon in high-energy heavy-ion collisions},  Phys.
  Rev. Lett. {\bf 77} (1996) 231--234
  [\href{http://arXiv.org/abs/hep-ph/9605213}{{\tt hep-ph/9605213}}].

\bibitem{Arleo:2004xj}
F.~Arleo, P.~Aurenche, Z.~Belghobsi and J.-P. Guillet, {\it Photon tagged
  correlations in heavy ion collisions},  JHEP {\bf 11} (2004) 009
  [\href{http://arXiv.org/abs/hep-ph/0410088}{{\tt hep-ph/0410088}}].

\bibitem{Adcox:2004mh}
{\bf PHENIX} Collaboration, K.~Adcox {\em et~al.}, {\it Formation of dense
  partonic matter in relativistic nucleus nucleus collisions at {RHIC}:
  Experimental evaluation by the {PHENIX} collaboration},  Nucl. Phys. {\bf
  A757} (2005) 184--283 [\href{http://arXiv.org/abs/nucl-ex/0410003}{{\tt
  nucl-ex/0410003}}].

\bibitem{Adams:2005dq}
{\bf STAR} Collaboration, J.~Adams {\em et~al.}, {\it Experimental and
  theoretical challenges in the search for the quark gluon plasma: The {STAR}
  collaboration's critical assessment of the evidence from {RHIC} collisions},
  Nucl. Phys. {\bf A757} (2005) 102--183
  [\href{http://arXiv.org/abs/nucl-ex/0501009}{{\tt nucl-ex/0501009}}].

\bibitem{Eskola:2005tx}
K.~J. Eskola, H.~Honkanen, H.~Niemi, P.~V. Ruuskanen and S.~S. R{\"a}s{\"a}nen,
  {\it Transverse spectra of hadrons in central {A} {A} collisions at {RHIC}
  and {LHC} from {pQCD} + saturation + hydrodynamics and from {pQCD} + energy
  losses},  talk given at {Quark Matter} 2005, {Budapest} ({Hungary}), 4-9
  {August} 2005, \href{http://arXiv.org/abs/hep-ph/0510019}{{\tt
  hep-ph/0510019}}.

\bibitem{Shimomura:2005pc} {\bf {PHENIX}} Collaboration, M.~Shimomura,
{\it High $p_{_\perp}$ $\pi^0$, $\eta$, identified  and inclusive
charged hadron spectra from PHENIX}, talk given at {Quark Matter}
2005, {Budapest} ({Hungary}), 4-9 {August} 2005,
\href{http://arXiv.org/abs/nucl-ex/0510023}{{\tt nucl-ex/0510023}}.

\bibitem{Vitev:2002pf}
I.~Vitev and M.~Gyulassy, {\it High {$p_{_\perp}$} tomography of d + {Au} and
  {Au} + {Au} at {SPS}, {RHIC}, and {LHC}},  Phys. Rev. Lett. {\bf 89} (2002)
  252301 [\href{http://arXiv.org/abs/hep-ph/0209161}{{\tt hep-ph/0209161}}].

\bibitem{Eskola:2004cr}
K.~J. Eskola, H.~Honkanen, C.~A. Salgado and U.~A. Wiedemann, {\it The
  fragility of high $p_{\perp}$ hadron spectra as a hard probe},  Nucl. Phys.
  {\bf A747} (2005) 511--529 [\href{http://arXiv.org/abs/hep-ph/0406319}{{\tt
  hep-ph/0406319}}].

\bibitem{Dainese:2004te}
A.~Dainese, C.~Loizides and G.~Paic, {\it Leading-particle suppression in high
  energy nucleus nucleus collisions},  Eur. Phys. J. {\bf C38} (2005) 461--474
  [\href{http://arXiv.org/abs/hep-ph/0406201}{{\tt hep-ph/0406201}}].

\bibitem{Turbide:2005fk}
S.~Turbide, C.~Gale, S.~Jeon and G.~D. Moore, {\it Energy loss of leading
  hadrons and direct photon production in evolving quark-gluon plasma},  Phys.
  Rev. {\bf C72} (2005) 014906 [\href{http://arXiv.org/abs/hep-ph/0502248}{{\tt
  hep-ph/0502248}}].

\bibitem{Borghini:2005em}
N.~Borghini and U.~A. Wiedemann, {\it Distorting the hump-backed plateau of
  jets with dense {QCD} matter},
  \href{http://arXiv.org/abs/hep-ph/0506218}{{\tt hep-ph/0506218}}.

\bibitem{Baier:1997kr}
R.~Baier, Y.~L. Dokshitzer, A.~H. Mueller, S.~Peign{\'e} and D.~Schiff, {\it
  Radiative energy loss of high energy quarks and gluons in a finite-volume
  quark-gluon plasma},  Nucl. Phys. {\bf B483} (1997) 291--320
  [\href{http://arXiv.org/abs/hep-ph/9607355}{{\tt hep-ph/9607355}}].

\bibitem{Baier:1997sk}
R.~Baier, Y.~L. Dokshitzer, A.~H. Mueller, S.~Peign{\'e} and D.~Schiff, {\it
  Radiative energy loss and {$p_{_\perp}$} broadening of high energy partons in
  nuclei},  Nucl. Phys. {\bf B484} (1997) 265--282
  [\href{http://arXiv.org/abs/hep-ph/9608322}{{\tt hep-ph/9608322}}].

\bibitem{Chiappetta:1996wp}
P.~Chiappetta, R.~Fergani and J.~P. Guillet, {\it Production of two large
  {$p_{_\perp}$} hadrons from hadronic collisions},  Z. Phys. {\bf C69} (1996)
  443--457.

\bibitem{Binoth:2002wa}
T.~Binoth, J.-P. Guillet, E.~Pilon and M.~Werlen, {\it A next-to-leading order
  study of photon pion and pion pair hadro-production in the light of the
  {Higgs} boson search at the {LHC}},  Eur. Phys. J. Direct {\bf C4} (2002) 7
  [\href{http://arXiv.org/abs/hep-ph/0203064}{{\tt hep-ph/0203064}}].

\bibitem{Piller:1999wx}
G.~Piller and W.~Weise, {\it Nuclear deep-inelastic lepton scattering and
  coherence phenomena},  Phys. Rept. {\bf 330} (2000) 1--94
  [\href{http://arXiv.org/abs/hep-ph/9908230}{{\tt hep-ph/9908230}}].

\bibitem{Arneodo:1994wf}
M.~Arneodo, {\it Nuclear effects in structure functions},  Phys. Rept. {\bf
  240} (1994) 301--393.

\bibitem{Accardi:2003be}
A.~Accardi {\em et~al.}, {\it Hard probes in heavy ion collisions at the {LHC}:
  {PDFs}, shadowing and p {A} collisions},  Report of the ``{PDFs}, shadowing
  and p {A} collisions'' in the CERN Yellow Report on ``Hard Probes in Heavy
  Ion Collisions at the LHC'', CERN--2004--009
  [\href{http://arXiv.org/abs/hep-ph/0308248}{{\tt hep-ph/0308248}}].

\bibitem{Eskola:1998df}
K.~J. Eskola, V.~J. Kolhinen and C.~A. Salgado, {\it The scale dependent
  nuclear effects in parton distributions for practical applications},  Eur.
  Phys. J. {\bf C9} (1999) 61--68
  [\href{http://arXiv.org/abs/hep-ph/9807297}{{\tt hep-ph/9807297}}].

\bibitem{Armesto:2003bq}
N.~Armesto and C.~A. Salgado, {\it Gluon distributions in nuclei at small
  {$x$}: Guidance from different models},
  \href{http://arXiv.org/abs/hep-ph/0301200}{{\tt hep-ph/0301200}}.

\bibitem{Baier:2001yt}
R.~Baier, Y.~L. Dokshitzer, A.~H. Mueller and D.~Schiff, {\it Quenching of
  hadron spectra in media},  JHEP {\bf 09} (2001) 033
  [\href{http://arXiv.org/abs/hep-ph/0106347}{{\tt hep-ph/0106347}}].

\bibitem{Arleo:2002kh}
F.~Arleo, {\it Tomography of cold and hot {QCD} matter: Tools and diagnosis},
  JHEP {\bf 11} (2002) 044 [\href{http://arXiv.org/abs/hep-ph/0210104}{{\tt
  hep-ph/0210104}}].

\bibitem{Salgado:2002cd}
C.~A. Salgado and U.~A. Wiedemann, {\it A dynamical scaling law for jet
  tomography},  Phys. Rev. Lett. {\bf 89} (2002) 092303
  [\href{http://arXiv.org/abs/hep-ph/0204221}{{\tt hep-ph/0204221}}].

\bibitem{Ashman:1991cx}
{\bf EMC} Collaboration, J.~Ashman {\em et~al.}, {\it Comparison of forward
  hadrons produced in muon interactions on nuclear targets and deuterium},  Z.
  Phys. {\bf C52} (1991) 1--12.

\bibitem{Airapetian:2003mi}
{\bf HERMES} Collaboration, A.~Airapetian {\em et~al.}, {\it Quark
  fragmentation to {$\pi^\pm$}, {$\pi^0$}, {$K^\pm$}, {$p$} and {$\bar{p}$} in
  the nuclear environment},  Phys. Lett. {\bf B577} (2003) 37--46
  [\href{http://arXiv.org/abs/hep-ex/0307023}{{\tt hep-ex/0307023}}].

\bibitem{Arleo:2003jz}
F.~Arleo, {\it Quenching of hadron spectra in {DIS} on nuclear targets},  Eur.
  Phys. J. {\bf C30} (2003) 213--221
  [\href{http://arXiv.org/abs/hep-ph/0306235}{{\tt hep-ph/0306235}}].

\bibitem{Pumplin:2002vw}
J.~Pumplin {\em et~al.}, {\it New generation of parton distributions with
  uncertainties from global {QCD} analysis},  JHEP {\bf 07} (2002) 012
  [\href{http://arXiv.org/abs/hep-ph/0201195}{{\tt hep-ph/0201195}}].

\bibitem{Kniehl:2000fe}
B.~A. Kniehl, G.~Kramer and B.~Potter, {\it Fragmentation functions for pions,
  kaons, and protons at next-to-leading order},  Nucl. Phys. {\bf B582} (2000)
  514--536 [\href{http://arXiv.org/abs/hep-ph/0010289}{{\tt hep-ph/0010289}}].

\bibitem{Bourhis:1997yu}
L.~Bourhis, M.~Fontannaz and J.-P. Guillet, {\it Quark and gluon fragmentation
  functions into photons},  Eur. Phys. J. {\bf C2} (1998) 529--537
  [\href{http://arXiv.org/abs/hep-ph/9704447}{{\tt hep-ph/9704447}}].

\bibitem{Bourhis:2000gs}
L.~Bourhis, M.~Fontannaz, J.-P. Guillet and M.~Werlen, {\it Next-to-leading
  order determination of fragmentation functions},  Eur. Phys. J. {\bf C19}
  (2001) 89--98 [\href{http://arXiv.org/abs/hep-ph/0009101}{{\tt
  hep-ph/0009101}}].

\bibitem{Fontannaz:2006pc}
P. Aurenche, M.~Fontannaz, J.-P. Guillet, E. Pilon, M. Werlen, in preparation.

\bibitem{Adler:2003pb}
{\bf {PHENIX}} Collaboration, S.~S. Adler {\em et~al.}, {\it Mid-rapidity
  neutral pion production in proton proton collisions at {$\sqrt{s}$} = 200
  {GeV}},  Phys. Rev. Lett. {\bf 91} (2003) 241803
  [\href{http://arXiv.org/abs/hep-ex/0304038}{{\tt hep-ex/0304038}}].

\bibitem{Okada:2005in}
{\bf {PHENIX}} Collaboration, K.~Okada, {\it Measurement of prompt photon in
  {$\sqrt{s}$}~=~200~{GeV} p p collisions},  talk given at SPIN2004, Trieste
  (Italy), 10-16 October 2004 [\href{http://arXiv.org/abs/hep-ex/0501066}{{\tt
  hep-ex/0501066}}].

\bibitem{Frantz:2004gg}
{\bf {PHENIX}} Collaboration, J.~Frantz, {\it {PHENIX} direct photons in 200
  {GeV} p + p and {Au} + {Au} collisions},
  \href{http://arXiv.org/abs/nucl-ex/0404006}{{\tt nucl-ex/0404006}}.

\bibitem{Isobe:2005pc}
{\bf PHENIX} Collaboration, T.~Isobe, {\it Neutral pion production in
  {$\sqrt{s}$}=200~{GeV} {Au} + {Au} collisions at {RHIC}-{PHENIX}
  collaboration},  shown at the {DNP/JPS} {Joint Fall Meeting}, Kapalua (Maui),
  Hawaii (USA), 18-22 {September} 2005.

\bibitem{Adler:2005ig}
{\bf {PHENIX}} Collaboration, S.~S. Adler {\em et~al.}, {\it Centrality
  dependence of direct photon production in {$\sqrt{s}$} = 200 {GeV} {Au} +
  {Au} collisions},  Phys. Rev. Lett. {\bf 94} (2005) 232301
  [\href{http://arXiv.org/abs/nucl-ex/0503003}{{\tt nucl-ex/0503003}}].

\bibitem{Jeon:2002dv}
S.~Jeon, J.~Jalilian-Marian and I.~Sarcevic, {\it The origin of
  large-{$p_{\perp}$} {$\pi^0$} suppression at {RHIC}},  Phys. Lett. {\bf B562}
  (2003) 45--50 [\href{http://arXiv.org/abs/nucl-th/0208012}{{\tt
  nucl-th/0208012}}].

\bibitem{Ferbel:1984ef}
T.~Ferbel and W.~R. Molzon, {\it Direct photon production in high-energy
  collisions},  Rev. Mod. Phys. {\bf 56} (1984) 181.

\bibitem{Baier:2002tc}
R.~Baier, {\it Jet quenching},  Nucl. Phys. {\bf A715} (2003) 209--218
  [\href{http://arXiv.org/abs/hep-ph/0209038}{{\tt hep-ph/0209038}}].

\bibitem{Arleo:2002ph}
F.~Arleo, {\it Constraints on quark energy loss from {Drell-Yan} data},  Phys.
  Lett. {\bf B532} (2002) 231 [\href{http://arXiv.org/abs/hep-ph/0201066}{{\tt
  hep-ph/0201066}}].

\bibitem{Back:2004je}
B.~B. Back {\em et~al.}, {\it The {PHOBOS} perspective on discoveries at
  {RHIC}},  Nucl. Phys. {\bf A757} (2005) 28--101
  [\href{http://arXiv.org/abs/nucl-ex/0410022}{{\tt nucl-ex/0410022}}].

\bibitem{Pisarski:1993rf}
R.~D. Pisarski, {\it Damping rates for moving particles in hot {QCD}},  Phys.
  Rev. {\bf D47} (1993) 5589--5600.

\bibitem{Peshier:2004bv}
A.~Peshier, {\it Hard gluon damping in hot {QCD}},  Phys. Rev. {\bf D70} (2004)
  034016 [\href{http://arXiv.org/abs/hep-ph/0403225}{{\tt hep-ph/0403225}}].

\bibitem{Zakharov:1996fv}
B.~G. Zakharov, {\it Fully quantum treatment of the {Landau-Pomeranchuk-Migdal}
  effect in {QED} and {QCD}},  JETP Lett. {\bf 63} (1996) 952--957
  [\href{http://arXiv.org/abs/hep-ph/9607440}{{\tt hep-ph/9607440}}].

\bibitem{Zakharov:1997uu}
B.~G. Zakharov, {\it Radiative energy loss of high energy quarks in finite-size
  nuclear matter and quark-gluon plasma},  JETP Lett. {\bf 65} (1997) 615--620
  [\href{http://arXiv.org/abs/hep-ph/9704255}{{\tt hep-ph/9704255}}].

\bibitem{Zakharov:2004bi}
B.~G. Zakharov, {\it Induced photon emission from quark jets in
  ultrarelativistic heavy-ion collisions},  JETP Lett. {\bf 80} (2004) 1--6
  [\href{http://arXiv.org/abs/hep-ph/0405101}{{\tt hep-ph/0405101}}].

\bibitem{Fries:2002kt}
R.~J. Fries, B.~M{\"u}ller and D.~K. Srivastava, {\it High energy photons from
  passage of jets through quark gluon plasma},  Phys. Rev. Lett. {\bf 90}
  (2003) 132301 [\href{http://arXiv.org/abs/nucl-th/0208001}{{\tt
  nucl-th/0208001}}].

\bibitem{d'Enterria:2005vz}
D.~d'Enterria and D.~Peressounko, {\it Probing the {QCD} equation of state with
  thermal photons in nucleus nucleus collisions at {RHIC}},
  \href{http://arXiv.org/abs/nucl-th/0503054}{{\tt nucl-th/0503054}}.

\bibitem{Huovinen:2001wx}
P.~Huovinen, P.~V. Ruuskanen and S.~S. R{\"a}s{\"a}nen, {\it Photon emission in
  heavy ion collisions at the {CERN} {SPS}},  Phys. Lett. {\bf B535} (2002)
  109--116 [\href{http://arXiv.org/abs/nucl-th/0111052}{{\tt
  nucl-th/0111052}}].

\bibitem{Gelis:2004ep}
F.~Gelis, H.~Niemi, P.~V. Ruuskanen and S.~S. R{\"a}s{\"a}nen, {\it Photon
  production from non-equilibrium {QGP} in heavy ion collisions},  J. Phys.
  {\bf G30} (2004) S1031--S1036
  [\href{http://arXiv.org/abs/nucl-th/0403040}{{\tt nucl-th/0403040}}].

\bibitem{Ahrens:2005pc} L. Ahrens {\it et al.}, {\it Luminosity
increases in Au--Au operation in RHIC}, talk given at EPAC2004,
Lucerne (Switzerland), 5-9 Jul 2004, EPAC-2004-MOPLT165.

\end{thebibliography}
\end{document}